\documentclass[preprint2]{aastex}

\usepackage{multirow}
\usepackage{amsmath}
\usepackage{float}
\usepackage{tabularx}
\usepackage{color}
\usepackage{subfigure}
\usepackage{graphicx}
\usepackage{ulem}
\usepackage{enumerate}

\newcommand{\datx}{x}
\newcommand{\daty}{y}
\newcommand{\fpsf}{f}
\newcommand{\fspec}{g}

\newcommand{\data}{\mbox{data}}
\newcommand{\genpara}{\psi}
\newcommand{\p}{\mu}
\newcommand{\n}{n}
\newcommand{\w}{w}
\newcommand{\pw}{\lambda}

\newcommand{\nk}{K}
\newcommand{\vk}{k}
\newcommand{\nktrue}{K_{\text{true}}}
\newcommand{\pk}{\kappa}  

\newcommand{\full}{F}
\newcommand{\gm}{\gamma}
\newcommand{\cpara}{\theta}
\newcommand{\cparas}{\Theta}

\newcommand{\spcparas}{\Theta^{\text{sp}}}

\newcommand{\extcparas}{\Theta^{\text{ext}}}
\newcommand{\extcpara}{\theta^{\text{ext}}}
\newcommand{\epa}{\alpha}

\newcommand{\mcestfk}{\hat{\w}_2^F}
\newcommand{\en}{E}
\newcommand{\alloc}{s}
\newcommand{\enc}{E}
\newcommand{\emax}{E_{\max}}
\newcommand{\emin}{E_{\min}}
\newcommand{\sig}{\sigma}
\newcommand{\rint}{r}

\newcommand{\sint}{m}
\newcommand{\psr}{q}
\newcommand{\bint}{b}
\newcommand{\polarr}{d}
\newcommand{\polart}{\omega}

\newcommand{\polarz}{d_0}
\newcommand{\cartx}{x}
\newcommand{\carty}{y}
\newcommand{\kingp}{\eta}
\newcommand{\post}{p}
\newcommand{\prior}{p}
\newcommand{\ewt}{\pi}
\newcommand{\ew}{\rho}
\newcommand{\Lfull}{L^{\text{full}}}
\newcommand{\Lsp}{L^{\text{sp}}}
\newcommand{\Lext}{L^{\text{ext}}}
\newcommand{\naive}{na\"{i}ve}

\newcommand{\coledits}{\color{black}}

\shorttitle{Overlapping Sources}
\shortauthors{Jones et al.}

\begin{document}

\slugcomment{\bf \today}

\title{Disentangling Overlapping Astronomical Sources using Spatial and Spectral Information}

\author{David E. Jones}
\affil{Statistics Department, Harvard University, 1 Oxford Street,
    Cambridge, MA 02138; djones@fas.harvard.edu.}
\author{Vinay L. Kashyap }
\affil{Harvard-Smithsonian Center for Astrophysics, 60 Garden Street, Cambridge, MA 02138.}

\author{David A. van Dyk}
\affil{Imperial College London, Statistics Section, Department of Mathematics, 180 Queen's Gate, London, UK SW7 2AZ.}

\begin{abstract}
We present a powerful new algorithm that combines  both spatial information (event locations and the point spread function) and spectral information (photon energies) to separate photons from overlapping sources. We use Bayesian statistical methods to simultaneously infer the number of overlapping sources, to probabilistically separate the photons among the sources, and to fit the parameters describing the individual sources. Using the Bayesian joint posterior distribution, we are able to coherently quantify the uncertainties associated with all these parameters. The advantages of combining spatial and spectral information are demonstrated through a simulation study. The utility of the approach is then illustrated by analysis of observations of FK Aqr and FL Aqr with the XMM-{\sl Newton} Observatory and the central region of the Orion Nebula Cluster with the {\sl Chandra} X-ray Observatory.
\end{abstract}


\section{Introduction}

When two or more sources are situated close enough to each other that there is a substantial overlap of their Point Spread Functions (PSFs), they pose a many-fold problem to astronomical analysis. The first is to recognize that there is an overlap, the second is to determine the number of distinct sources that are involved, the third is to measure their relative intensities, and the fourth is to separate them sufficiently to be able to carry out useful secondary analyses like spectral fitting and variability analysis. These problems are especially complicated for high-energy photon detectors, since the data are firmly in the Poisson regime, background is often a significant component of the data, and the simplifying approximations of a Gaussian process are usually inapplicable. Many researchers have considered the simpler problem of a single source contaminated by background in the low counts regime (e.g., \citealt{kraft1991}, \citealt{loredo1992}, \citealt{high_energy_bayes}, \citealt{park2006}, \citealt{weisskopf2007}, \citealt{laird2009}, \citealt{knoetig2014}, \citealt{primini2014}), and have generally found that Poisson-likelihood based Bayesian techniques are well suited to address this category of problems.

However, in the case of multiple sources, progress has been slow, and the choices limited. One could construct approximate measures of intensities of the component sources in the Gaussian regime via matrix inversion (\citealt{kashyap1994}), or choose to minimize contamination by limiting the sizes of the apertures to cover only the cores of the PSFs (\citealt{broos2010}), or carry out full-fledged 2-D spatial modeling. All these are approximate or computationally intensive solutions. An important advance was made recently by \citet{primini2014}, who developed a fully Bayesian aperture photometry method that simultaneously models the intensities of the overlapping sources and the intensity of the background. Their method can be applied to any counts image with multiple overlapping sources, with a practical computational limit of up to five sources. Despite this, most of the problems listed above are still extant.

\begin{table*}[t]
\centering
\caption{Symbols used in this work. Notation used only in a single section is defined where it appears and is not included in this table.\vspace{0.25cm}}
\resizebox{\linewidth}{!}{
\begin{tabular}{|l|l|}
\hline
Symbol & Definition \\
\hline
$(\datx_i,\daty_i)$ & Location of photon $i$ on the detector\\
$\enc_i$ & Energy (PI channel) of photon $i$ \\
$\p_j$ & True location of source $j$ (2-D coordinates)\\
$\fpsf_{\p_j}$ & Point Spread Function centered at $\p_j$\\
$\epa_j$ & Spectral shape parameter for source $j$ (full model) \\
$\gm_j$ & Spectral mean parameter for source $j$ (full model)\\
$\epa_{jl}$ & Spectral shape parameter $l$ for source $j$ (extended full model) \\
$\gm_{jl}$ & Spectral mean parameter $l$ for source $j$ (extended full model)\\
$\pi_{jl}$ & Weight in $[0,1]$ of {\sl gamma} component $l$ in source $j$ spectral model (extended full model)\\
$\emin,\emax$ & Minimum and maximum detected energy (PI channel)\\
$\w_j$ & Relative intensity of source $j$ ($j=0$ for background)\\
$\nk$ & True number of sources ($\nktrue$ for emphasis)\\
$\vk$ & A possible value of $\nk$ \\
$\pk$ & Prior mean of $\nk$ \\
$\alloc_i$ & The true source of photon $i$ (takes the values $0,\dots,\nk$, with $0$ indicating background)\\
$\n_j$ & True number of photons detected from source $j$ ($j=0$ for background)\\
$\n$ & Total number of photons detected i.e. $\sum_{j=0}^\nk \n_j$\\
$\cpara_j$ & Full model parameters for source $j$ i.e. $\{\w_j,\p_j,\epa_j,\gm_j\}$\\
$\cparas_\nk$ & All full model source specific parameters i.e. $\{\cpara_0,\dots,\cpara_\nk\}$ where $\cpara_0=\w_0$\\
$\Lfull,\Lsp,\Lext$ & Likelihood function of the full, spatial-only, and extended full models, respectively \\
$\genpara^{(t)}$ & The value of generic parameter $\genpara$ in iteration $t$ of the algorithm \\
$\datx,\daty,\enc,\alloc$ & Vectors of the corresponding photon specific variables (see earlier table entries)\\
$I_A$ & Indicator function equal to 1 if the event $A$ occurs (e.g. $\nk=3$) and 0 otherwise\\
\hline
\end{tabular}
}
\end{table*}

Typically, X-ray data are collected as lists of events, with each event tagged by its location on the detector, its energy,\footnote{The detector records the pulse height amplitude (PHA), which is roughly proportional to the energy of the incoming photon. These values are often reported as pulse-invariant (PI) gain-corrected PHAs. The distribution of PI for a photon at a given energy is encoded in the detector's Redistribution Matrix File (RMF). In the following, we use ``energy" as a synonym for this recorded PI, and clarify only if there is any ambiguity.} and its arrival time. Binning the positions into images causes a loss of information that could be alleviated by carrying out the analysis on the unbinned event lists. In such a case, it becomes feasible to disentangle individual events and allocate them probabilistically to the several sources that comprise the dataset. In the following, we describe an algorithm that directly addresses three of the four problems listed above: it dynamically determines the number of overlapping sources, measures their intensities, and pools individual events into clusters for which follow-up spectral analysis can be carried out. {\coledits There are related approaches for longer wavelength data originating from an unknown number of sources, for example, \citealt{brewer2013} and \citealt{safarzadeh2014}. The former uses Gaussian process models to identify stellar oscillation modes, and the latter uses simulated Herschel images based on Hubble data to investigate a disentangling method. The principal difference between these methods and our approach is that they conduct analysis at the pixel level, whereas we probabilistically assign individual photons to sources, a key distinction when analyzing low-count X-ray data.}

\subsection{Statistical Approach}

Here we use finite mixture distributions to model several overlapping sources of photons in a high-energy image. Finite mixture distributions are a useful class of statistical models for data that are drawn from a mixture of several subpopulations; these models are finite in that the (possibly unknown) number of subpopulations is a finite positive integer. (See \citealt{2004finite} and \citealt{titterington1985} for comprehensive discussion of finite mixture distributions, and, for example, \citealt{mixture} for a previous application in astronomy.)  We take a Bayesian perspective that allows joint inference for the parameters that describe the photon sources (e.g., their number, intensities and locations), the basic shape of their spectra, and the probability that any particular photon originated from each source, given its recorded location and energy.

Performing inference jointly on the image and spectra  improves the precision of the fitted parameters, and also provides more coherent measures of uncertainty than would be available if the spatial and spectral  data were analyzed separately. Furthermore, unlike other methods for overlapping sources, our approach quantifies uncertainty about the number of sources. Whether we are ultimately interested in spatial or spectral aspects of sources, identifying the correct number of sources is clearly fundamental. Consequently, a coherent measure of the uncertainty associated with the fitted number of sources is critical to the appropriate interpretation of the fitted parameters of the individual sources.

In some applications inference for the number of sources may seem unnecessary because the sources are clearly identifiable. For instance, the XMM-{\sl Newton} observation of FK Aqr and FL Aqr analyzed in Section \ref{sec:xmm} has relatively weak background noise, and the sources overlap only moderately. In such cases, the main advantage of the proposed method is that it precisely quantifies the uncertainties associated with the positions, intensities and spectral shapes of the sources. As already mentioned, finite mixture analysis also yields, for each observed photon, the probability that it originated from each inferred source (or the background). In this way we do not deterministically assign photons to sources, but rather properly assess the uncertainty of their origins. This is in contrast to other methods, such as those based on source regions, which deterministically assign photons to nearby sources, and therefore do not properly quantify uncertainties in fitted source parameters.

{\coledits There is a potential for overfitting in finite mixture models if the number of sources is unknown. This is mitigated when substantial prior information regarding the shape of the PSF or the number of sources is available, or both. In practice,} we have detailed information about the PSF, and hence know exactly what the distribution of the recorded photon locations should be for each source. (For point sources this is trivial, but even for extended sources one can easily convolve the source model with the PSF.) Even if the PSF varies across the field, the shape of the photon scatter is completely determined by the location of the source. {\coledits With this complete knowledge of the PSF, there is only a small risk of overfitting, even with limited prior information regarding the number of sources and their spectral shapes. Indeed, our results do not strongly depend on the choice of prior distribution for the number of sources (see Section \ref{sec:ksense}).}

Our method is designed for analyzing images composed of an unknown number of point sources that are contaminated with background. However, it can be applied to extended sources, with some modifications to account for spatial variations in intensity and spectra. We also mention that the success of our method depends partly on our ability to use spectra to distinguish point sources from the background, which is possible because a typical X-ray point source {\coledits spectrum} is more peaked than the background. Because of this, we are able to use basic models that capture the rough spectral shape in order to exploit spectral information whilst conserving computational resources. In the X-ray band, this approach offers substantial improvements  over analyses using only spatial data  without the cost of precisely modeling the spectra.  However, the utility of the method in other wavelength bands will depend somewhat on the nature of the spectra typical of those bands.

The remainder of the paper is organized into seven sections. Section \ref{sec:data_model} develops the statistical model for isolated sources in the context of high-energy datasets, and describes how these models are combined in the case of multiple sources. Section \ref{sec:example} uses a motivating example to illustrate the method and the benefits of incorporating spectral models for the sources. The beginning of Section \ref{sec:bayesian_analysis} gives a brief review of Bayesian inference. The remainder of Section \ref{sec:bayesian_analysis} describes the details of the proposed Bayesian analysis and computational approach. Section \ref{sec:sims} presents two simulation studies. The first illustrates that inference for the number of sources is insensitive to the choice of prior distribution, and the second more thoroughly studies the advantages of using the spectral data. Sections \ref{sec:xmm} and \ref{sec:chandra} present the results of our analysis of observations from the XMM-{\sl Newton} and {\sl Chandra} X-ray observatories. The {\sl XMM} observation is of the apparent visual binary FK Aqr and FL Aqr, and the {\sl Chandra} observation is of approximately 14 sources from near the center of the Orion nebula. We summarize in Section \ref{sec:summary} and computational details are in the appendices. {\coledits Our {\it Bayesian Separation of Close Sources (BASCS)} software is available on GitHub at \texttt{https://github.com/astrostat/BASCS}.}

\section{Data and Statistical Models}
\label{sec:data_model}

\subsection{Structure of the Data}
\label{sec:data}

High-energy detectors record directional coordinates $(\datx_i, \daty_i)$ and energy $\enc_i$ for each detected photon, where $i=1,\dots ,n$ indexes the photons. As mentioned, in practice, the PI channel is used to quantify energy. We denote the full set of spatial and spectral information for $\n$ detected photons by $(\datx,\daty,\enc)$. These observed quantities are subject to the effects of the PSF and the spectral Redistribution Matrix Function (RMF). We explicitly account for PSF effects in our model, but model the {\it observed} spectra, the convolution of the source spectra and the RMF. This strategy does not allow us to fit source spectral models, but does allow us to leverage spectral data  to separate the sources. Even though all the attributes are recorded digitally and are binned quantities, we treat them as continuous variables for simplicity, since this binning is at scales that heavily over-sample the PSF.

Each photon is assumed to originate from one of several point sources or the background, but its exact origin is unknown. Furthermore,  the number of point sources contributing photons to the data, their locations, intensities, and spectral distributions are all unknown. We assume background is distributed uniformly across the image, its strength and spectral distribution are often not known.

\subsection{Prototype Model for a Single Source}
\label{sec:one_model}

To introduce notation and our model in the simplest case, we first suppose that the data consist of photons from a single source, with no background contamination. Statistical models specify a distribution for the observed data conditional on a number of typically unknown parameters; we discuss parameter fitting Section \ref{sec:inference}. In the current case, given the unknown position  of the source, the detected photons are assumed to be dispersed according to a PSF. That is,
\begin{eqnarray}
(\datx_i,\daty_i)|\p \sim \mbox{PSF centered at }\p
\end{eqnarray}
for $i = 1,\dots,\n$, where $\p=(\p_x,\p_y)$ is the unknown position of the source\footnote{The notation $x|z\sim F$ means that, the variable $x$ has the distribution denoted by $F$ if $z$ is fixed and known, and  we say that $x$, given $z$, follows the distribution $F$. Throughout, when we use this notation we mean that repeated realizations of $x$ are independent given $z$.}. We use the same 2-D King profile\footnote{The \texttt{beta2d} model in CIAO/{\sl Sherpa}.} in all the simulations and data analyses presented, see \citet{read2010} and \citet{king}. The King profile density, shown in Figure \ref{fig:king} in Appendix C, has heavy tails and is essentially a bivariate Cauchy distribution. Specific parameter values are detailed in Appendix C. More generally, although our method assumes that the PSF is known given $\p$, it may vary with $\p$. Furthermore, the PSF may be any function which can be quickly evaluated analytically or numerically. Even in cases where computationally expensive evaluations are required our method is feasible if the PSF is first tabulated.

An important feature of our overall approach is that it also utilizes the spectral data to better assess the likely origin of each photon (when background or more than one source is present). With this end in mind, we propose a simple and computationally practical model for the basic shape of the source spectrum. In particular, we model photon energies using a {\sl gamma} distribution,\footnote{A standard parameterization of the {\sl gamma}$(\alpha,\beta)$ distribution yields the density $f(x)=\frac{\beta^\alpha}{\Gamma(\alpha)}x^{\alpha-1}e^{-\beta x}= \frac{\epa^\epa}{\gm^\epa\Gamma(\epa)}x^{\alpha-1} e^{-\frac{\epa}{\gm} x}, x>0$. Here $\alpha$ and $\beta=\epa/\gm$ are the shape and rate parameters, respectively.}
\begin{eqnarray}
\enc_{i}|\epa,\gm &\sim& \mbox{{\sl gamma}}(\epa,\epa/\gm)
\label{eqn:spectral1}
\end{eqnarray}
for $i = 1,\dots,\n$. Here, $\epa$ and $\gm$ are the unknown shape and mean parameters used to describe the basic spectral distribution.\footnote{We parameterize the {\sl gamma} distribution using the shape and mean, instead of the shape and rate, for interpretability and because computationally it is best to avoid rates, which in our applications tend to be close to the parameter space boundary at zero.} The {\sl gamma} distribution allows flexible modeling of positive quantities with right skewed distributions.\footnote{Indeed, the Exponential and Chi-squared distributions are special cases, and a {\sl gamma} can also closely resemble a (truncated) Gaussian distribution.} We emphasize that we aim to summarize the essential shape of the spectral distribution, rather than to model the details of emission lines and other spectral features. This is practical because for high-energy missions, the effective areas are typically small at low and high energies, with a broad peak in the middle; the resulting counts spectrum is reasonably modeled by a single- or double-component {\sl gamma} distribution (particularly since we ignore the RMF). Our goal is to identify sources and divide photons among them, not to carry out detailed spectral analysis. However, our algorithm allows for complex spectral models to be built in if necessary. In addition, and computationally more feasible, once the {\sl gamma} model has fulfilled its role in separating sources, a more sophisticated spectral model may then be used to draw scientific conclusions about the spectral distributions of the disentangled sources. This final stage will be discussed in Section \ref{sec:secondary_analysis}.

\subsection{Prototype Model for Multiple Sources}
\label{sec:more_model}

In practice there are multiple sources and background contamination, hence we introduce a {\it finite mixture model}. Let $\nk$ be a  parameter denoting the number of sources and $\p = (\p_1,\dots,\p_\nk)$ be a $2\times \nk$ matrix giving the source positions i.e. $\p_j= (\p_{jx},\p_{jy})$, for $j=1,\dots,\nk$. If we knew the origin of every photon then, we could model the spatial and spectral data associated with each point source as we did in Section \ref{sec:one_model}.  We thus introduce a new variable $\alloc_i$ which indicates the source number associated with photon $i$. Each $\alloc_i$ takes on a value between $1$ and $\nk$, and we let $\alloc$ denote the vector $(\alloc_1,\dots,\alloc_\n)$. Note that $\alloc_i$ is never actually observed and thus is a {\it latent variable}. A latent variable is essentially an unknown parameter which is useful for modeling, but may not be of direct interest in itself.  Here, we have introduced $\alloc_i$ to simplify the model and to facilitate the algorithms used for inference, {\coledits which are described} in Section \ref{sec:inference}.

As a parameter, $\alloc_i$ is also conditioned on in our spatial model, which now becomes
\begin{eqnarray}
(\datx_i,\daty_i)|(\p,\alloc_i=j) \sim \mbox{PSF centered at }\p_j
\label{eqn:xymodel}
\end{eqnarray}
for $i = 1,\dots,\n$. As an unknown parameter, $\alloc_i$, plays a role similar to $\mu$; it is ``given" in Equation \ref{eqn:xymodel}. The spectral model can also be straightforwardly generalized to the multiple source case. We have
\begin{eqnarray}
\en_{i}|(\epa_j,\gm_j,\alloc_i=j) \sim \mbox{{\sl gamma}}(\epa_j,\epa_j/\gm_j)
\label{eqn:spectral}
\end{eqnarray}
for $i = 1,\dots,\n$, where the parameters $\epa_j$ and $\gm_j$ usually differ among the sources.

In addition to point sources, we must model the background. To this end we extend the set of possible values of $\alloc_i$ to include $0$. Throughout, symbols indexed by $0$ refer to the background. We assume that photons originating from the background are uniformly distributed across the image,
\begin{eqnarray}
(\datx_i,\daty_i)|(\p,\alloc_i=0) \sim \mbox{Uniform}
\label{eqn:xybmodel}
\end{eqnarray}
for $i$ such that $\alloc_i=0$. Instrument effects may cause the background to be non-uniform, and a refinement would be to model such effects.

The background spectrum is also assumed to be flat over the energy range of the source spectra. That is, it is assumed to have a uniform distribution on $(\emin,\emax)$, where $\emin$ and $\emax$ are the minimum and maximum photon energy observed. This is a good approximation because the background spectrum is expected to be less peaked than that of a point source.

So far we have not considered the intensities of the different sources and the background. Naturally there should be a parameter for each source, and one for the background, to specify the intensities. Let $\n_j$ denote the number of photons originating from source $j$, for $j=0,\dots,\nk$ (with zero denoting the background), mathematically,\footnote{$I$ is an indicator function that is zero if its argument is false and one otherwise.} $\n_j = \sum_{i=1}^\n I_{\{\alloc_i=j\}}$. We can realistically model $\n_j$ as a Poisson variable with some mean $m_j$, for $j = 0,\dots,\nk$. Because these Poisson means vary with exposure time, however, the relative intensities, $\w_j = m_j/\sum m_j$, are of more direct interest. Writing $\w = (\w_0,\dots,\w_\nk)$, and given $\n$, the Poisson model for $(\n_0,\dots,\n_\nk)$ yields a Multinomial model,\footnote{The Multinomial distribution assigns the probability $(n!/(n_0!\cdots n_\nk!)) \prod \w_0^{n_0}\cdots \w_\nk^{n_\nk}$  to the allocation given by $(n_0,\dots,n_\nk)$ of $n = \sum_{i=0}^\nk n_i$ objects into $\nk+1$ categories.}
\begin{eqnarray}
(\n_0,\dots,\n_\nk)|\w,\n,\nk\sim \mbox{Multinomal}(\n;\w),
\label{eqn:mult}
\end{eqnarray}
where  $\sum_{j=0}^\nk \w_j=1$. Under this parameterization, the relative strengths of the sources and background can be succinctly expressed by the vector $\w = (\w_0,\dots,\w_\nk)$ without further reference to $\n$. Accordingly, all inference is performed given $n$, because its value tells us nothing about the number of sources or their parameters.

To complete our introduction of the model we derive the likelihood function, which is the probability of the data expressed as a function of the parameters. The likelihood tells us what values of the parameters are supported by the data and is a key component for principled statistical inference. Let $\mathcal{I}_j$ be the set of photons originating from source $j$ (including $j=0$) and let $\mathcal{I}$ be the entire collection of observed photons.\footnote{Mathematically, $\mathcal{I}_j$ is the set of photon indices associated with source $j$, that is, $\mathcal{I}_j = \{i:\alloc_i=j\}$, for $j=0,\dots,\nk$, and $\mathcal{I}=\bigcup_{j=0}^\nk\mathcal{I}_j=\{1,\dots,\n\}$.} Also, denote the value of the PSF centered at $\p$ and evaluated at $(\datx,\daty)$ by $\fpsf_\p(x,y)$. Lastly, here and throughout, we let $\cpara_j=\{\w_j,\p_j,\epa_j,\gm_j\}$ denote the parameters associated with source $j$, for $j=1,\dots,\nk$. Similarly, for the background, we let $\cpara_0=\w_0$. We let $\cparas_\nk$ denote all the source (and background) specific parameters i.e. $\cparas_\nk=\{\cpara_0,\dots,\cpara_\nk\}$. The remaining parameters are $\nk$ and $\alloc$. As already discussed, we treat $\n$ as fixed, and impose the constraint that the likelihood is zero unless $\sum_{j=0}^\nk \n_j=n$. Combining the different parts of the model yields the {\it full model} likelihood
\begin{eqnarray}
\Lfull_n(\cparas_\nk,\nk) \equiv p(\datx,\daty,\en|\cparas_\nk,\nk,\alloc,n) \nonumber\\
\propto \prod_{i \in \mathcal{I}/\mathcal{I}_0}\fpsf_{\p_{\alloc_i}}(\datx_i,\daty_i)\fspec_{\epa_{\alloc_i},\gm_{\alloc_i}}(\en_i),
\label{eqn:likelihood}
\end{eqnarray}
where
\begin{eqnarray}
\fspec_{\epa_{\alloc_i},\gm_{\alloc_i}}(\en_i)=\frac{\epa_{\alloc_i}^{\epa_{\alloc_i}}}{\gm_{\alloc_i}^{\epa_{\alloc_i}}
\Gamma(\epa_{\alloc_i})}\en_i^{\alpha_{\alloc_i}-1}e^{-\epa_{\alloc_i}\en_i/\gm_{\alloc_i}}.
\end{eqnarray}
The maximum energy $\emax$ and the image area are assumed to be known quantities, rather than parameters to be inferred. They are therefore omitted from the likelihood, as are all terms not involving the parameters. In later sections, we compare analyses under the full model to analyses under the {\it spatial-only model} that does not use the spectral information. The likelihood of the spatial-only model is
\begin{eqnarray}
\Lsp_n(\spcparas_\nk,\nk) &\equiv& p(\datx,\daty|\spcparas_\nk,\nk,\alloc,n) \nonumber\\
&\propto& \prod_{i \in \mathcal{I}/\mathcal{I}_0} \fpsf_{\p_{\alloc_i}}(\datx_i,\daty_i).
\label{eqn:spatial_likelihood}
\end{eqnarray}
The notation $\spcparas_\nk=\{\w_0,\dots,\w_\nk;\p_1,\dots,\p_\nk\}$ represents the set of spatial parameters.  Note that, although $\w$ does not explicitly appear in either likelihood, the data does nevertheless constrain $\w$ in both cases. In particular, the likelihoods indicate probable values of $\alloc$ which in turn indicate probable values of $\w$. Conceptually, our method is to apply Bayes rule, briefly reviewed in Section \ref{sec:basic_bayes}, to the likelihoods displayed in Equations \ref{eqn:likelihood} and \ref{eqn:spatial_likelihood} to yield a distribution summarizing our knowledge of the parameters given the data, i.e. the joint posterior distribution.

\subsection{Extensions of the spectral model}
\label{sec:spectral_extensions}

\begin{figure}[t]
\begin{center}
\includegraphics[trim  = 0mm 10mm 0mm 10mm, width=0.45\textwidth]{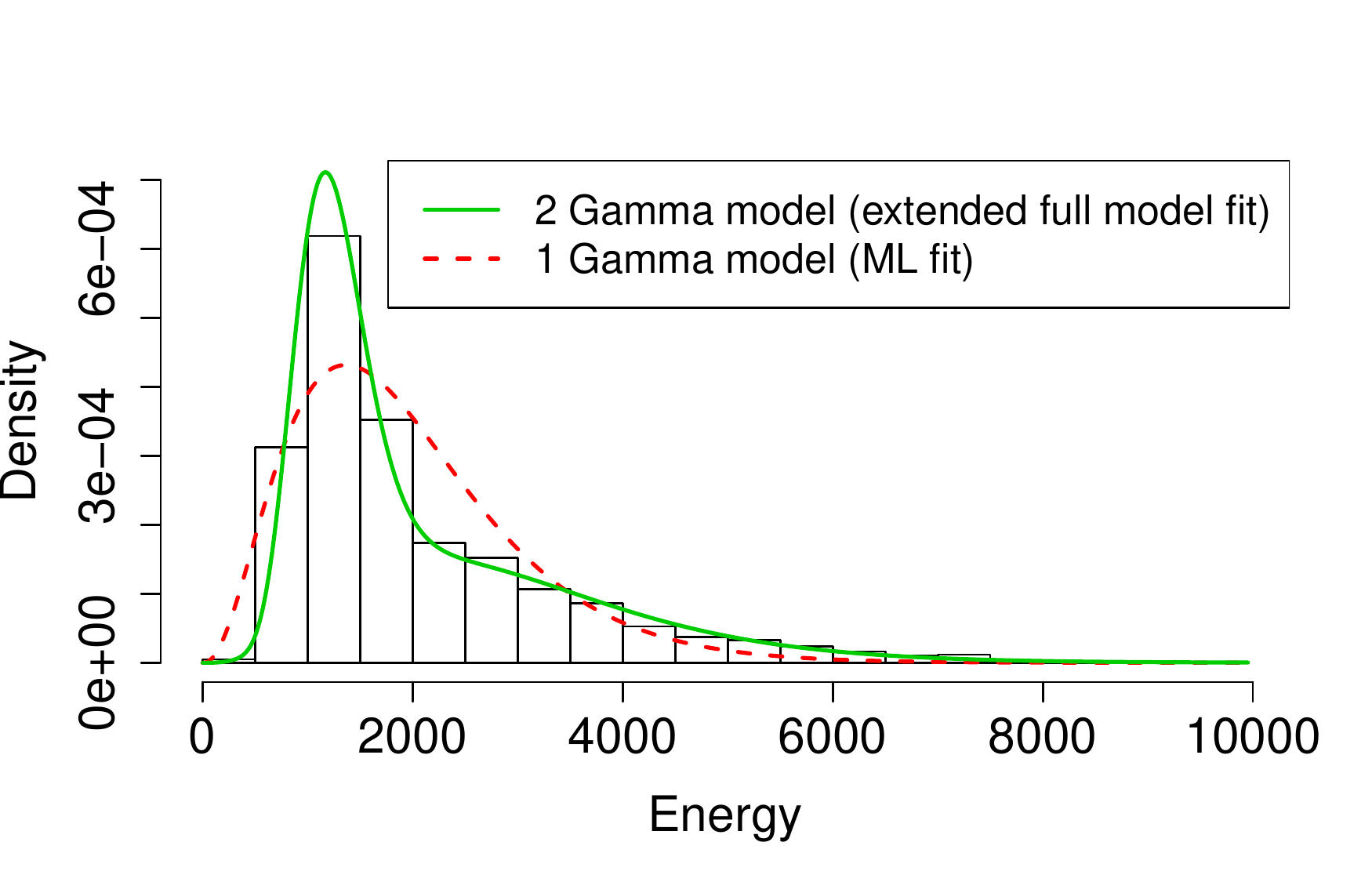}
\caption{Fitting {\sl gamma} distributions to a counts spectrum. The histogram shows the observed spectrum of the brightest of the Chandra sources in the Orion field in Section \ref{sec:secondary_analysis} (from one iteration of our algorithm; see Section \ref{sec:inference}), and the curves show {\sl gamma} model fits. The solid line (green) is the extended full model fit of the two-{\sl gamma} spectral model and the dashed line (red) is the maximum likelihood fit of the one-{\sl gamma} model. \label{fig:gam2spectral}}
  \end{center}
\end{figure}

In some situations the {\sl gamma} spectral model given by Equation \ref{eqn:spectral} is not sufficiently flexible to capture the spectral shape of the observed sources. For example, Figure \ref{fig:gam2spectral} shows the observed spectrum of the brightest source in the {\sl Chandra} observation analysed in Section \ref{sec:chandra}. In particular, the histogram shows the spectrum using one likely assignment of photons produced during the iterations of our algorithm (see Section \ref{sec:inference}). The dashed red curve shows the maximum likelihood fit of the {\sl gamma} distribution to the observed spectrum. The {\sl gamma} does not fit the distribution closely. This causes a problem because inference based on the (misspecified) {\sl gamma} spectral model will suggest there are two sources instead of one in order to better capture the spectral distribution of the source.

To solve this problem, we use a mixture of two {\sl gamma} distributions for a more general spectral model. That is,
\begin{eqnarray}
\en_{i}|(\epa_{j1},\epa_{j2},\gm_{j1},\gm_{j2},\ewt_{j1},\ewt_{j2},\alloc_i=j)  \nonumber \\
\sim \sum_{l=1}^2\ewt_{jl}\mbox{{\sl gamma}}\left(\epa_{jl},\frac{\epa_{jl}}{\gm_{jl}}\right),
\label{eqn:spectral_2gam}
\end{eqnarray}
for $i = 1,\dots,\n$, where the parameters $\ewt_j$ and $\ewt_{j2}=1-\ewt_{j1}$ are the weights of the two {\sl gamma} components. When this two {\sl gamma} mixture spectral model is substituted for the one {\sl gamma} spectral model in Equation \ref{eqn:likelihood} we obtain the following {\it extended full model} likelihood
\begin{eqnarray}
\Lext_n(\extcparas_\nk,\nk)\equiv p(\datx,\daty,\en|\extcparas_\nk,\nk,\alloc,n) \nonumber \\
\propto \prod_{i \in \mathcal{I}/\mathcal{I}_0}
\left(\fpsf_{\p_{\alloc_i}}(\datx_i,\daty_i)
 \sum_{l=1}^2\ewt_{\alloc_{i}l} \fspec_{\epa_{\alloc_{il}},\gm_{\alloc_{il}}}(\en_i)\right).
\label{eqn:likelihood_extended}
\end{eqnarray}
The notation $\extcparas_\nk$ denotes $\{\extcpara_0,\dots,\extcpara_\nk\}$, where $\extcpara_j=\{\w_j,\p_j,\epa_{j1},\epa_{j2},\gm_{j1},\gm_{j2},\ewt_{j1},\ewt_{j2}\}$ gives the parameters associated with source $j$, for $j=1,\dots,\nk$, and $\extcpara_0=\cpara_0$. The solid green curve in Figure \ref{fig:gam2spectral} shows the extended full model fit of the {\sl gamma} mixture spectral model. In this example, the mixture of {\sl gammas} quite closely fits the observed spectrum and generally there did not appear to be unwarranted splitting of sources into two in our numerical studies using this model.

Even greater flexibility of the spectral model could be gained by considering a mixture of more than two {\sl gammas}, but this was not necessary in our numerical studies. For the {\sl XMM} data of Section \ref{sec:xmm}, the one-{\sl gamma} spectral model is sufficient in that, for both of the sources, the maximum likelihood fit of the one-{\sl gamma} and the two-{\sl gamma} models resulted in essentially identical fits when using a feasible allocation of photons. In the interest of simplicity, we only use the extended full model when necessary (i.e., in Section \ref{sec:chandra}), and elsewhere use the full model given in Equation \ref{eqn:likelihood}.

\begin{figure*}[t]
\includegraphics[trim  = 0mm 5mm 0mm 0mm, width=1\textwidth]{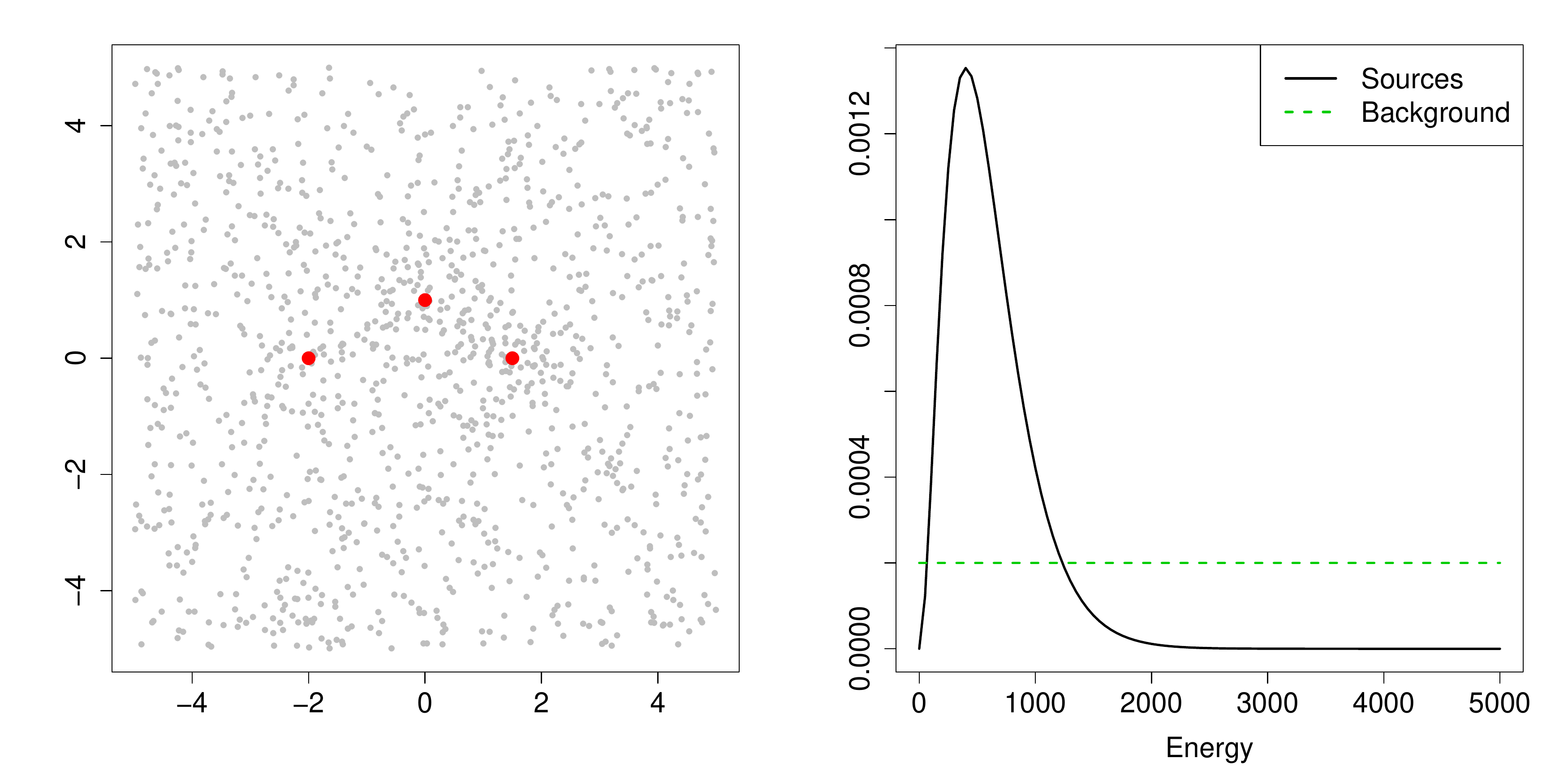}
\caption{Illustrative simulation setup. Locations of three weak sources are shown as red dots over a scatter plot (left), as also are the adopted counts spectra of the sources and the background (right).\label{fig:example}}
\end{figure*}

\subsubsection{Detecting spectral model inadequacy}

A natural question is how one should decide if the source spectral model is inadequate for our purpose of allocating photons among the different sources (and background). There are two potential indications of spectral model misspecification. Firstly, analysis may tend to divide bright sources into two. In particular, when the algorithm (see Section \ref{sec:inference}) finds many instances of sources very close together this indicates that the spectral model is probably not adequate.\footnote{\coledits{Misspecification of the PSF, and specifically under-estimation of its width, could have a similar effect.}} A second indication of inadequacy of the spectral model comes from considering inference under the spatial-only model. We can inspect the empirical distribution of the photons assigned to a source in iterations of the spatial-only algorithm.  If this empirical distribution differs substantially from a {\sl gamma} distribution then it is unlikely that the one-{\sl gamma} spectral model is sufficiently flexible. Clearly, looking at the empirical spectral distribution of a source under the spatial-only model is only reliable if we can accurately assign photons based on spatial data alone. Thus, when possible it is best to select a bright source which is relatively isolated. In the presence of uncertainty about the shape of the spectral distributions to expect then it is usually sensible to use a mixture of at least two {\sl gammas} (or perform analysis several times using mixtures of different numbers of {\sl gammas}). In the presence of uncertainty about the shape of the spectral distributions to expect then it is usually sensible to use a mixture of
at least two {\sl gammas} (or perform analysis several times using mixtures of different numbers of {\sl gammas}). One should be cautious of using a spectral model that is too complicated\footnote{{\coledits We can avoid an overly complicated model by imposing parametric constraints or utilizing substantial prior information to be sure only scientifically plausible spectral shapes are allowed.}} because overfitting may decrease the benefits of modeling the spectral data.

\section{Illustrative example}
\label{sec:example}

To motivate our method we present a simple simulated data example that illustrates the potential gains made possible by using the full model instead of the spatial-only model. We emphasize that this is a walk-through, designed to clarify the conceptual foundations of the method. A detailed description of our method is in Section \ref{sec:bayesian_analysis}. The simulated data consist of the spatial and spectral details of photons detected from three weak sources contaminated with background. The spatial data and the spectral distributions used for simulation are shown in Figure \ref{fig:example}. The background average is 10 photons per unit square, and the numbers of photons from each source are drawn from Poisson distributions with means 100, 50 and 25, respectively. Thus, the background is very strong and contributes about 85\% of the photons over the entire image, and about 40\%, 53\%, and 66\% respectively in the three source regions.  The true source positions are $(1.5,0)$, $(0,1)$, and $(-2,0)$, and their source regions are approximately circles of radius 1. All three sources have the same PSF, the 2-D profile density shown in Figure \ref{fig:king} in Appendix C. The source spectral data is drawn from a {\sl gamma} distribution plotted in Figure \ref{fig:example} (mean parameter 600 and shape parameter 3). In this simple illustration, all the sources have the same theoretical spectral distribution; however, this is not assumed in the fitted model, which is based on the likelihood in Equation \ref{eqn:likelihood}. The theoretical background spectrum was uniform on $(0,5000)$.

\begin{figure*}[p]
\includegraphics[trim  = 0mm 10mm 0mm 0mm, width=1\textwidth]{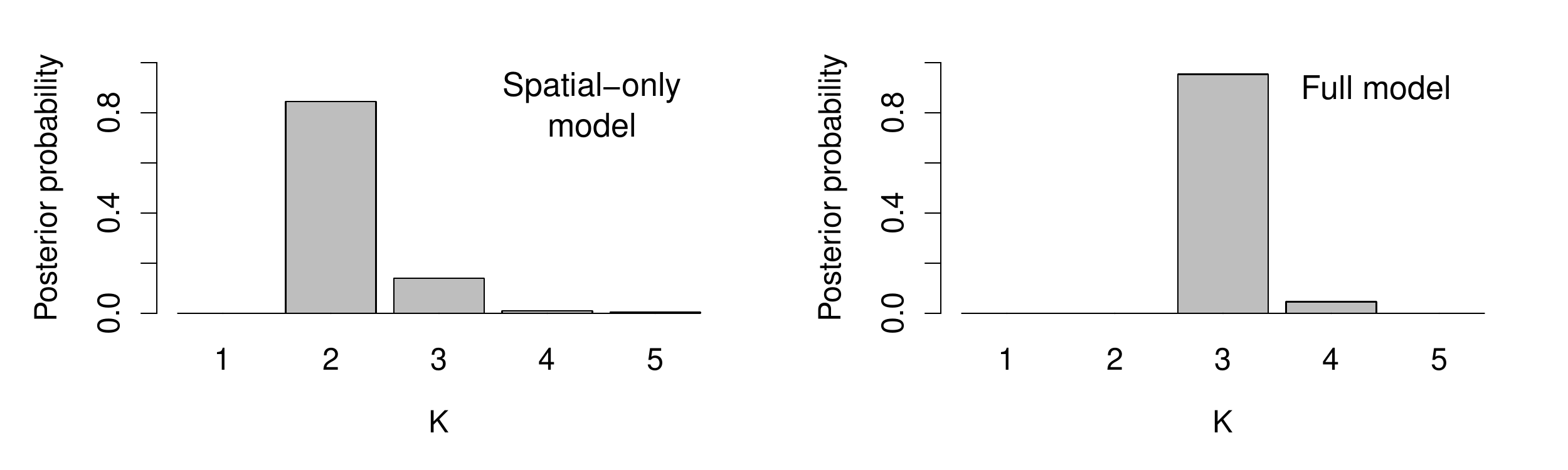}
\caption{Probability distribution of the  number of sources based on the spatial-only model (left) and the full model (right). In this simulation, the true value is $K=3$. \label{fig:3postk}}
\end{figure*}

\begin{table*}[p]
\caption{Fitted parameters under the full and spatial-only models. The columns in bold give the fits that would likely be relied upon in practice for the two models. The intervals in parentheses indicate the 16\% and 84\% posterior quantiles, i.e., Bayesian $1\sigma$ equivalent intervals. \label{tab:parasweak}\vspace{0.25cm}}
\centering
\resizebox{0.80\linewidth}{!}{
\begin{tabular}{l|c|rl|rl|rl}
  \hline
 & Truth &   \multicolumn{2}{c|}{Full model} &  \multicolumn{4}{c}{Spatial-only model}  \\
  \hline
$\vk$ & 3 & \textbf{3} & &\textbf{2}& & 3  &  \\
$P(\nk=\vk|\mbox{data})$ &-- &\textbf{0.95} & & \textbf{0.85}&  &  0.14 &   \\
\hline
 $\mu_{1x}$  & 1.5 & \textbf{1.51} &(1.41,1.61) & \textbf{1.43} &(1.27,1.58)  & 1.44 &(1.29,1.59)  \\
  $\mu_{1y}$ & 0 & \textbf{$-$0.01} &($-$0.10,0.09) & \textbf{0.04} &($-$0.08,0.17) & 0.02 &($-$0.10,0.14)  \\
 $\mu_{2x}$ & 0 & \textbf{$-$0.08} &($-$0.20,0.04) & \textbf{$-$0.09} &($-$0.28,0.12) & $-$0.03 &($-$0.22,0.15)  \\
$\mu_{2y}$ & 1  & \textbf{1.11} &(1.00,1.23) & \textbf{0.96} &(0.80,1.13) & 0.99 &(0.84,1.15) \\
  $\mu_{3x}$   & $-$2 & \textbf{$-$1.96} &($-$2.17,$-$1.76) &--&-- & $-$1.37 &($-$2.40,0.35) \\
 $\mu_{3y}$   & 0 & \textbf{0.06} &($-$0.15,0.27) &--&-- & $-$0.24 &($-$1.44,0.75) \\
  $w_1$  & 0.083 & \textbf{0.068} &(0.057,0.078) & \textbf{0.063} &(0.049,0.076) & 0.062 &(0.049,0.076) \\
   $w_2$ & 0.058 & \textbf{0.064} &(0.053,0.076) & \textbf{0.055} &(0.041,0.068)  & 0.052 &(0.039,0.066) \\
   $w_3$  & 0.033  & \textbf{0.028} &(0.019,0.036) &--&--& 0.017 &(0.003,0.030)\\
   $w_0$ & 0.826  & \textbf{0.841} &(0.826,0.855) & \textbf{0.883} &(0.866,0.900)& 0.868 &(0.848,0.887)\\
   $\gm_1$  & 600 & \textbf{536} &(478,592) &--&-- &--&--\\
  $\gm_2$ & 600 & \textbf{735} &(646,820) &--&--&--&--\\
  $\gm_3$ & 600 & \textbf{634} &(397,826) &--&-- &--&--\\
  $\epa_1$ & 3 & \textbf{3.92} &(2.89,4.97) &--&--&--&--\\
  $\epa_2$ & 3 & \textbf{2.94} &(2.18,3.69) &--&--&--&--\\
  $\epa_3$& 3 & \textbf{2.76} &(1.62,3.82) &--&--&--&--\\
   \hline
\end{tabular}}
\end{table*}

\begin{table*}[p]
   \caption{Photon allocation proportions for the spatial-only and full models. \label{tab:assignment}\vspace{0.25cm}}
  \resizebox{ \linewidth}{!}{%
    \begin{tabular}{l|c|cccc||cccc}
    \hline
    \multirow{3}[0]{*}{Source (true intensity)} & \multirow{3}[0]{*}{\parbox{2.5cm}{No. Photons in simulation}} &\multicolumn{8}{c}{Average allocation probabilities} \\ \cline{3-10}
    & & \multicolumn{4}{c||}{Spatial-only model} & \multicolumn{4}{c}{Full model}\\
          &       & Background   & Right& Middle& Left& Background   & Right& Middle& Left  \\
          \hline
    Background (10/sq) &  1001 & 0.917 & 0.037 & 0.033 & 0.013 & 0.940 & 0.022 & 0.026 & 0.012 \\
    Right (100) &84 & 0.566 & 0.354 & 0.068 & 0.012& 0.318 & 0.557 & 0.113 & 0.012 \\
    Middle (70) & 67 & 0.593 & 0.073 & 0.303 & 0.031 & 0.313 & 0.122 & 0.505 & 0.060 \\
        Left (40) & 42 & 0.800 & 0.034 & 0.071 & 0.095 & 0.431 & 0.066 & 0.145 & 0.358 \\
    \hline
    \end{tabular}%
    }
\end{table*}

We fit both the spatial-only and the full model to the simulated data. The resulting posterior probability distributions for $\nk$ are shown in Figure \ref{fig:3postk}. With the spatial data alone it is difficult to detect the faintest source, and consequently the most likely value of  $\nk$  is 2 rather than 3. The situation is much improved when we include the spectral data. The advantage of using the spectral information is due to a greater ability to distinguish the sources from the background, owing to the difference between the source spectra and the background spectrum.

Modeling the spectral data also improves estimation of the other parameters, even if we consider the fits based on $\nk=3$. (This is the correct value of $\nk$ and is identified by the full model but not the spatial-only model.) In Table \ref{tab:parasweak}, the first bold column and the last column (not bold) show a summary of the fitted parameters for $\nk=3$ under the full model and spatial-only model, respectively. When we consider $\nk=3$, the greatest gains of using the full model are in estimating the parameters of the faintest source because this source is the hardest to distinguish from the background when using only spatial data.

In practice, the advantage of using the spectral data for estimating the source parameters is greater than is apparent when we only consider $\nk=3$. When confronted with the summary of the fit of $\nk$ under the spatial-only model (given in the left panel of Figure \ref{fig:3postk}), a researcher is likely to rely on the parameter fits assuming $\nk=2$. Thus, it is fair to compare the $\nk=3$ fit under the full model with the $\nk=2$ fit under the spatial-only model (i.e., the bold columns in Table \ref{tab:parasweak}). The latter is clearly substantially worse than the former, because the faint source goes undetected and has no fitted parameters.

The improvement in separation of the sources (and background) can be further understood from Table \ref{tab:assignment}, which summarizes the probability that each photon originated from each source or the background, again under the optimistic assumption that $\nk=3$ (see Section \ref{sec:inference} for additional details). The rows of the table indicate the true photon origin, and the columns indicate the fitted origins. The table entries are the average probabilities, across photons, of the different fitted origins. Ideally the matrices would be identity matrices with `1's along the diagonal and `0's elsewhere, but because of the strength of the background many source photons are mixed up in the background. For example, for a photon originating from the leftmost source, the spatial-only model on average assigns probabilities of 0.095 and 0.800 that it originated from the correct source and the background respectively, reflecting the difficulty in detecting the location of this faint source. Under the full model, the average probability of correct assignment is increased to 0.358, a substantial improvement. Indeed, for each of the three sources, {\it nearly half as many photons are mixed up with the background under the full model}. Our improved ability to correctly assign photons under the full model (relative to the spatial-only model) naturally leads to improved estimation of the parameters of the faint source, as illustrated in Table \ref{tab:parasweak}. There is a similar effect for the other sources though it is less pronounced because, being brighter, they are easier to detect from the spatial data alone.

\section{Bayesian Model Fitting}
\label{sec:bayesian_analysis}

\subsection{Bayesian Inference}
\label{sec:basic_bayes}

The Bayesian perspective provides a coherent approach for combining all available information to infer the unknown model parameters $\cparas_\nk$, $\nk$, and $\alloc$. Firstly, our knowledge (or lack of knowledge) as to the likely values of the parameters before seeing the current data is quantified using a {\it prior distribution}. Once the data are observed, Bayes' Theorem allows us to combine the likelihood and the prior distribution to yield the {\it posterior distribution} of the parameters. Recall, the likelihood is the probability of the data given the parameters. The posterior distribution expresses our updated knowledge of the parameters after seeing the data. Bayes' Theorem states that, for generic data and parameter vector $\genpara$, the posterior distribution is
\begin{eqnarray}
\label{eqn:bayes_theorem}
\post(\genpara|\data) = \frac{p(\data|\genpara)\prior(\genpara)}{p(\data)},
\end{eqnarray}
where $p(\data|\genpara)$ is the likelihood function and $p(\genpara)$ is the prior distribution. The denominator $p(\data)$ is simply a normalizing constant which ensures the posterior integrates to one.  In our case, the data is $(\datx,\daty,\enc)$ and under the full model $\genpara=\{\cparas_\nk,\nk,\alloc\}$ so
\begin{eqnarray}
\label{eqn:our_bayes_theorem}
\post(\cparas_\nk,\nk,\alloc|\datx,\daty,\en) \qquad\qquad\qquad\qquad\nonumber\\
= \frac{p(\datx,\daty,\en|\cparas_\nk,\nk,\alloc)\prior(\cparas_\nk,\nk,\alloc)}{p(\datx,\daty,\en)}.
\end{eqnarray}
Here, all probabilities are conditional on $n$ but this is suppressed. The likelihood $p(\datx,\daty,\en|\cparas_\nk,\nk,\alloc)$ is given in Equation \ref{eqn:likelihood}, and the prior distribution $\prior(\cparas_\nk,\nk,\alloc)$ is described in Section \ref{sec:prior}. Referring back to the illustrative example in Section \ref{sec:example}, the marginal posterior distribution of $\nk$,
\begin{eqnarray}
p(\nk|\datx,\daty,\en) \quad\qquad\qquad\qquad\qquad\qquad\nonumber\\
= \sum_{\alloc}\int p(\nk,\alloc,\cparas_\nk | \datx,\daty,\en) d\cparas_\nk,
\label{eqn:kmarginal}
\end{eqnarray}
is displayed in Figure \ref{fig:3postk}. Given the number of unknown parameters, it is not possible to plot their joint posterior distribution, but we can derive and plot the marginal posterior distribution of any one parameter, as in Equation \ref{eqn:kmarginal} and Figure \ref{fig:3postk}.

\subsection{Completing the Model Formulation: Prior Distributions}
\label{sec:prior}

Following the Bayesian approach, we specify prior distributions for each of the unknown parameters. Firstly, the positions of the point sources are {\it a priori} assumed to be independently and uniformly distributed across the image. That is,
\begin{eqnarray}
\p_j \sim \mbox{Uniform}
\end{eqnarray}
for $j=1,\dots,\nk$. In principle, informative priors can be used if prior information on source locations is available. For example, we might set $\p_j \sim N(\p_{j0},\sigma_{j0}^2)$, where $(\p_{j0},\sigma_{j0})$, for $j=1,\dots,\nk$, specifies knowledge of the source locations.\footnote{The notation $N(\p_{j0},\sigma_{j0}^2)$ denotes a Gaussian distribution with mean $\p_{j0}$ and variance $\sigma_{j0}^2$.}

Next, the vector $\w$, that specifies relative intensities, is given a Dirichlet\footnote{The Dirichlet density is
$f(p_0,\dots,p_\nk)=\left(\Gamma(\sum_{i=0}^\nk \lambda_i)/\prod_{i=0}^\nk \Gamma(\lambda_i)\right) \prod_{i=0}^\nk p_i^{\lambda_i-1}$,  for all $p_i$ such that $\sum_{i=0}^\nk p_i =1$ and $p_i \geq 0$ for $i=0,\dots,\nk$, and is zero otherwise. Here,  $(\lambda_0,\dots,\lambda_\nk)$ is a  parameter, and $\Gamma$ is the gamma function.} prior distribution with parameter $(\pw,\dots,\pw)$. A Dirichlet random variable is a probability vector, i.e., it is a vector with non-negative entries that sum to one. We set $\pw=1$ throughout. This choice is uniform on the probability vector, but very slightly favors sources of equal size. Indeed, setting $\pw=1$ means the Dirichlet prior has as much information as a single photon count added to each source (including a single count added to the background).\footnote{Suppose the source counts are observed to be $(\n_0,\dots,\n_\nk)$ and follow a Multinomial distribution with probability vector $\w$. Then, assuming {\it a priori} $\w \sim \mbox{Dirichlet}(\pw_0,\dots,\pw_\nk)$, it can be shown that $\w|(\n_0,\dots,\n_\nk)\sim \mbox{Dirichlet}(\n_0+\pw_0,\dots,\n_\nk+\pw_\nk)$. Because $\pw_j$ is treated just like $\n_j$ in this posterior distribution, $\pw_j$ can be viewed as a ``prior count" and we say the Dirichlet prior is as informative as $\lambda_j$ counts added to source $j$, for $j=0,\dots,\nk$.} Regarding the realized vector of source and background counts $(\n_0,\dots,\n_\nk)$, recall that Equation \ref{eqn:mult} specifies a Multinomial distribution for $(\n_0,\dots,\n_\nk)$, given $\w$, $\n$, and $\nk$. Since $(\n_0,\dots,\n_\nk)$ is a function of the parameter (or latent variable) $\alloc$, Equation \ref{eqn:mult} is effectively a prior distribution for $\alloc$.\footnote{The parameter $\w$ is called a {\it hyper-parameter} because it appears in the prior distribution of $\alloc$ but is itself of interest and thus has its own prior distribution.}

External information about the number of sources is amalgamated into a prior for $\nk$, which we assume to be Poisson with mean parameter $\pk$.\footnote{{\coledits While other priors for $\nk$ are possible, the Poisson is simple and only moderately informative. Indeed, the equality of mean and variance captures the typical level of prior information we expect, e.g., if we suspect 10 sources, then an analysis yielding between 8 and 12
sources would seem quite reasonable, but we are unlikely to consider, say, 100 sources as a realistic possibility. Even less informative priors may sometimes be desirable, but it generally makes sense to use any reliable prior information that is available to guard against model misspecification. (Prior information about $\nk$ also helps our algorithm to converge slightly more quickly.)}} Under the Poisson prior, the fitted value of $\nk$ is relatively robust to the choice of $\pk$ because the PSF is completely specified.\footnote{If the PSF were not fully specified, it would be difficult to distinguish a few sources with a wide PSF from many sources with a narrow PSF. Thus, the fitted number of components of a general finite mixture model can be quite sensitive to the choice of prior on this parameter. Accounting for misspecified PSFs or uncertainties in their calibration is beyond the scope of this work (see \citealt{calibration} and \citealt{xu2014} for possible strategies).} Indeed, we show in Section \ref{sec:ksense} that the posterior mode for the number of sources may correctly identify the true value of $\nk$, even when $\pk$ is quite different from $\nk$. {\coledits Therefore, in practice it is adequate to use the Poisson prior for $\nk$ with $\pk$ set to any reasonable guess of the number of sources.}

To complete the model specification, we must assign prior distributions for the source spectral distribution parameters $\epa_j$ and $\gm_j$, for $j=1,\dots,\nk$. Typically there is sufficient data to overwhelm these prior distributions. Thus, we are not overly concerned with the exact form of these priors. For concreteness, however, we mention that one set of priors we use is $\epa_j \sim \mbox{{\sl gamma}}(2,0.5)$ and $\gm_j \sim \mbox{Uniform}(\emin,\emax)$, for $j=1,\dots,\nk$, where $\emin$ is the minimum observed energy.\footnote{{\coledits More generally, if $\nk$ is large and some of the sources are faint, it may be beneficial to model the distribution of the spectral parameters across the sources. This strategy is known as hierarchical modeling and is known to have statistical advantages in terms of the estimates of the individual spectral parameters. Such hierarchical spectral structures are left as a topic for future work.}}

To summarize, our prior distribution for the full model parameters $\cparas_\nk$, $\nk$ and $\alloc$ is
\begin{eqnarray}
\label{eqn:prior}
\prior(\cparas_\nk,\nk,\alloc) = p(\p,\epa,\gm,\alloc|\nk,\w)p(\w|\nk)p(\nk) \propto \nonumber\\
\left(\prod_{j=0}^\nk{\epa_j}e^{-0.5\epa_j}\right)\prod_{j=0}^\nk\w_j^{\n_j}\left(\prod_{j=0}^\nk \w_j\right)^{\pw-1} \frac{\pk^\nk}{\nk!},
\end{eqnarray}
where $\p$, $\epa$ and $\gm$ denote $(\p_1,\dots,\p_\nk)$, $(\epa_1,\dots,\epa_\nk)$, and $(\gm_1,\dots,\gm_\nk)$, respectively. The second term on the second line of Equation \ref{eqn:prior} comes from the Multinomial prior distribution for $\alloc$. In the case of the extended full model given in Equation \ref{eqn:likelihood_extended}, the priors for $\epa_{jl},\gm_{jl}$, $l=1,2$, are the same as those for $\epa_j,\gm_j$, and the prior for $\ewt_{j1}$ is a Beta$(2,2)$ distribution,\footnote{For $\alpha,\beta > 0$, the $\mbox{Beta}(\alpha,\beta)$ distribution density is $f(x) = \left(\Gamma(\alpha+\beta)/\Gamma(\alpha)\Gamma(\beta)\right)x^{\alpha-1}(1-x)^{\beta-1}$ for $x\in[0,1]$, and is zero otherwise. Here, $\Gamma$ is the Gamma function. } for $j=1,\dots,\nk$. (No prior for $\ewt_{j2}$ is needed because this parameter is determined by $\ewt_{j1}$, for $j=1,\dots,\nk$.) The prior for the spatial model parameters is Equation \ref{eqn:prior}  without the first term.

\subsection{Statistical Computation and Model Fitting}
\label{sec:inference}

Given the likelihood in Equation \ref{eqn:likelihood} and the prior distribution in Equation \ref{eqn:prior}, we can apply Bayes' Theorem to obtain the posterior distribution of $\cparas_\nk$, $\nk$ and $\alloc$ (see Equation \ref{eqn:our_bayes_theorem}). The resulting posterior distribution is a complicated function, which we summarize by the low-dimensional marginal distributions as described in Section \ref{sec:basic_bayes} and their means and standard deviations. These summaries are used to estimate the model parameters and their error bars.

We accomplish the necessary numerical integration, e.g., as in Equation \ref{eqn:kmarginal}, using Monte Carlo methods, a cornerstone of statistical computing (\citealt{shao2000}, \citealt{liu2008}, \citealt{2011handbook}). The idea of Monte Carlo algorithms is to simulate values of the generic parameter $\genpara$ from the posterior distribution in Equation \ref{eqn:bayes_theorem} to obtain a {\it Monte Carlo sample} $\{\genpara^{(1)},\dots,\genpara^{(T)}\}$. For example, in Figure \ref{fig:3postk}, the height of the bin centered at $\vk$ is the proportion of the Monte Carlo draws with $\nk^{(t)}$ equal to $\vk$, i.e.,
\begin{eqnarray}
\label{eqn:mc_kest}
P(\nk=\vk|\datx,\daty,\en)\approx \frac{1}{T}\sum_{t=1}^T I_{\{\nk^{(t)}=k\}},
\end{eqnarray}
for $\vk=1,\dots,\nk$.

A somewhat unusual feature of our model is that the number of parameters is determined by the value of $\nk$, the unknown number of sources. This necessarily conditional structure means that it only makes sense to consider the posterior distributions of the other parameters for a given inferred value of $\nk$ (\citealt{park} discuss a somewhat similar conditional inference in the context of locating emission lines). For an illustration of why this is so, consider the intensity $\w_3$ of the `third' source in an image. The parameter $\w_3$ does not have the same interpretation when there are three sources versus four, because what is the `third' source in the first scenario may combine two sources from the latter scenario. In fact, for $\nk=2$ the parameter $\w_3$ does not even exist. In general, there is no clear relationship between the parameters under scenarios with different values of $\nk$.  This prevents us from considering the unconditional posterior distribution of, say, $\w_3$. Instead, we are interested in posterior summaries given a particular value of $\nk$, such as $\post(\w_3|\nk=k,\datx,\daty,\en)$. For example, the second row of  Table \ref{tab:parasweak} provides an estimate of the posterior mean of $\w_2$ conditional on $\nk=3$, under the full model,\footnote{The superscript $\full$ in Equation \ref{eqn:w2k_estimate} indicates that the Monte Carlo samples were drawn from the posterior derived under the full model.}
\begin{eqnarray}
\label{eqn:w2k_estimate}
\mcestfk(\vk) = \frac{\sum_{t=1}^T \w_2^{(t)}I_{\{\nk^{(t)}=k\}}}{\sum_{t=1}^TI_{\{\nk^{(t)}=k\}}}=0.080.
\end{eqnarray}
More generally, for each one-dimensional parameter $\tau$, we calculate the Monte Carlo estimate
\begin{eqnarray}
\label{eqn:generalk_estimate}
\hat{\tau}(\vk) = \frac{\sum_{t=1}^T \tau^{(t)}I_{\{\nk^{(t)}=k\}}}{\sum_{t=1}^TI_{\{\nk^{(t)}=k\}}}.
\end{eqnarray}
In practice, we choose a value of $\vk$ at which $\nk$ has relatively high posterior probability, such as the posterior mode, because otherwise the parameters estimated are unlikely to correspond to properties of real sources. (Indeed, our algorithm does not accurately estimate parameters under unlikely values of $\nk$.) We may decide to consider several different values if the posterior of $\nk$ is not concentrated on one value. This can be useful despite the fact that, as we have mentioned, the number and interpretation of the parameters is not consistent across values of $\nk$.

The most popular method for obtaining the Monte Carlo samples needed for estimates such as that in Equation \ref{eqn:w2k_estimate} is Markov chain Monte Carlo (MCMC). This an iterative algorithm in which we generate a new value of the parameters $\genpara^{(t)}$ at each iteration by drawing from a distribution $\mathcal{F}$ that only depends on $\genpara^{(t-1)}$ (and the data) and not earlier members of the Monte Carlo sample. Continuing for $T$ iterations we obtain a sample $\{\genpara^{(1)},\dots,\genpara^{(T)}\}$ of correlated parameter values, {\coledits which is usually called an MCMC chain.} Appropriate choice of $\mathcal{F}$ ensures that the sample mimics the posterior distribution in the sense that as $T \rightarrow \infty$ the sample empirical distribution approaches the posterior distribution. In implementation, a draw from an appropriate $\mathcal{F}$ is typically achieved through two steps: firstly a new value of the parameters $\genpara^*$ is proposed, and then this value is either accepted or rejected with some probability.\footnote{An appropriate choice of $\mathcal{F}$ and the corresponding rejection probability to use, to ensure convergence of the sample empirical distribution to the posterior, can be calculated by appealing to the `reversibility condition' (see texts on the theory of Markov chain convergence e.g. \citet{feller1968}).} The Metropolis-Hastings algorithm (\citealt{metropolis} and \citealt{hastings}) is an example of such an algorithm. The reader is referred to \citet{textbook} for details, including discussion of efficient choices of $\mathcal{F}$ and monitoring of convergence to the posterior distribution {\coledits (which is usually done by running multiple MCMC chains in parallel and checking that their behaviour is sufficiently similar based on some criterion).}

In standard MCMC algorithms the parameter space being explored is fixed throughout. In our context this means the number of sources would have to be known. We therefore turn to reversible jump Markov chain Monte Carlo (RJMCMC) algorithms (first introduced by \citealt{green}), which allow configurations with differing numbers of sources to be explored. {\coledits There have been a number of uses of RJMCMC in other astronomy contexts, for example, \citealt{umstatter2005}, \citealt{brewer2009}, \citealt{wandelt2013}, and \citealt{walmswell2013}.} In RJMCMC algorithms, so called `jump' steps update the value of $\nk$, the name referring to a jump between parameter spaces (or `models'). These steps are performed by drawing $\nk^{(t)}$ from a distribution only depending on $\genpara^{(t-1)}=(\cparas^{(t-1)},\nk^{(t-1)},\alloc^{(t-1)})$, in the same spirit as ordinary MCMC iterations. Feasible values of the parameters $\cparas^{(t)}$ and $\alloc^{(t)}$ must simultaneously be drawn because their dimension and interpretation change with $\nk$. It is this high dimensional sampling that makes RJMCMC challenging. In RJMCMC algorithms, $\nk^{(t)}$ is only allowed to differ from $\nk^{(t-1)}$ by at most one. This constraint facilitates the proposal of appropriate parameters $\cparas^{(t)}$ and $\alloc^{(t)}$; RJMCMC moves between configurations by splitting, combining, creating or destroying sources in the model. The standard RJMCMC algorithm for Gaussian mixtures was introduced in \citet{richard}, and \citet{gammamixtures} illustrated RJMCMC for {\sl gamma} mixtures. Our {\coledits BASCS software} essentially combines these two algorithms. Additional details are given in Appendices A and B. {\coledits For the analyses found in Sections \ref{sec:sims} and \ref{sec:chandra} we specify the number of iterations for which our RJMCMC algorithm was run (which depended on the observed convergence rate and run time). A single iteration of our RJMCMC algorithm consists of one proposal to change $\nk$ and ten MCMC updates of the other parameters, i.e., the number of MCMC iterations is ten times greater than the stated number of RJMCMC iterations. In Section \ref{sec:xmm} we fix $\nk$ and use MCMC, and thus directly specify the number of MCMC iterations. Our standard approach is to run ten RJMCMC (or MCMC) chains to allow monitoring of convergence, but for simplicity the final results are always computed using a single chain.}

As discussed in Section \ref{sec:prior}, having detailed information about the PSF means our estimates are insensitive to the prior on $\nk$ (see also Section \ref{sec:ksense}). Knowledge of the PSF also aids computation in that it limits the number of feasible configurations, meaning the RJMCMC algorithm does not have to jump across many values of $\nk$. This keeps the number of iterations until approximate convergence comparatively low. Thus, knowledge of the PSF means that, despite the difficulties {\coledits that are commonly thought to surround} mixture models fit with RJMCMC algorithms, our proposed approach is relatively stable and robust. Nonetheless, when the number of sources is clear, MCMC algorithms should be used because they are computationally preferable to RJMCMC algorithms (see Section \ref{sec:xmm} for an analysis using an MCMC algorithm). In particular, MCMC algorithms are faster per iteration and fewer iterations are needed to obtain enough samples for a given $\nk$ value of interest. One further challenge is moderate sensitivity to the spectral model, which is the reason why in some applications the {\sl gamma} spectral model must be replaced by the {\sl gamma} mixture spectral model introduced in Section \ref{sec:spectral_extensions}.

\section{Simulation studies}
\label{sec:sims}

Simulated data are used to assess two important aspects of our method: (i) the sensitivity of the fit for $\nk$ on its prior distribution; and (ii) the performance of the method under a range of different source and background parameters. In the second case, of particular interest is the comparison of inference for the parameters under the spatial-only model and full models (given in Equations \ref{eqn:likelihood} and \ref{eqn:spatial_likelihood}).

\subsection{Sensitivity to prior distribution on $\nk$}
\label{sec:ksense}

\begin{figure*}[t]
\includegraphics[width=1\textwidth]{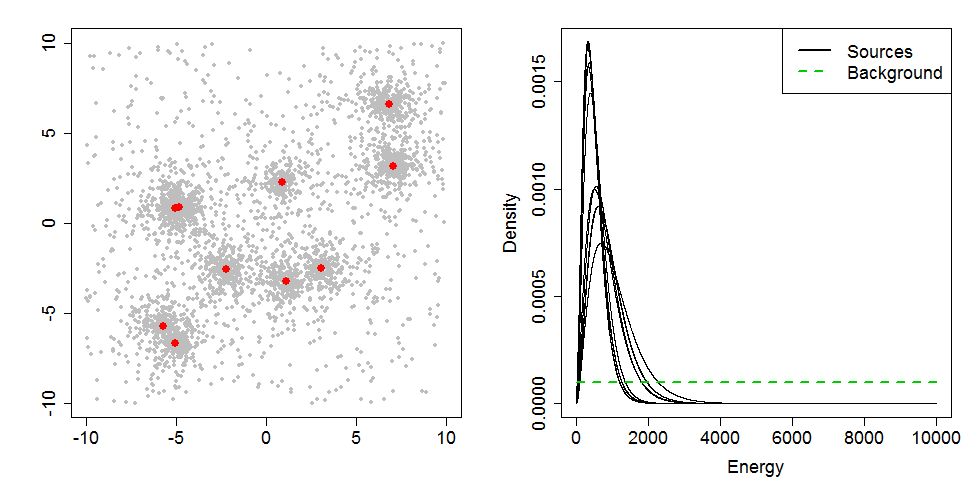}
\caption{Simulated dataset for the 10 source case. The simulated spatial counts distribution (left) and the
adopted spectra for each source and the background (right) are shown. The true locations of the 10 sources are marked by large (red) dots in the left plot.\label{fig:10examples}}
\end{figure*}

To illustrate robustness to the prior on $\nk$, we simulated data for a one-source ($\nktrue =1$) and a ten-source ($\nktrue =10$) reality and drew inference for the number of sources under three different settings of the prior mean $\pk$ ($1$, $3$, and $10$). Ten datasets were simulated under each reality, each consisting of images of 20 by 20 spatial units and spectral data (simulated under the single {\sl gamma} spectral model). We randomly placed the sources in the central 18 by 18 region of the image, avoiding the edges so that source photons are largely contained within the image. The mean number of photons $m_j$ from source $j$ was chosen randomly from the interval $100$ to $500$, for $j=1,\dots,\nk$. The mean total number of photons from the background in each dataset, $m_0$, was set to 400, an average of 1 photon per unit square. The number of photons from source $j$ (or the background) was then simulated from a Poisson distribution with mean $m_j$, for $j=0,\dots,\nk$. Spatial coordinates for the photons were chosen by sampling from the PSF (or the Uniform distribution in the case of the background). We used the King profile density for the PSF; the same PSF is used for analysis of the datasets in Section \ref{sec:xmm} and Section \ref{sec:chandra}.

To complete the datasets, we simulated spectral data under the single {\sl gamma} spectral model (and from a Uniform distribution in the case of the background). We drew the spectral distribution parameters $\epa_j$ (shape) and $\epa_j/\gm_j$ (rate), $j=1,\dots,\nk$, from the Gaussian distributions $N(3,0.2^2)$ and $N(0.005,0.001^2)$ respectively, truncating both distributions to be strictly positive. The resulting spectral parameters are similar to those fitted for the {\sl XMM} dataset in Section \ref{sec:xmm}. An example simulated dataset is shown in the left panel of Figure \ref{fig:10examples}. The right panel shows the true spectral distributions for the same dataset.

\begin{figure*}[p]
\includegraphics[width=1\textwidth]{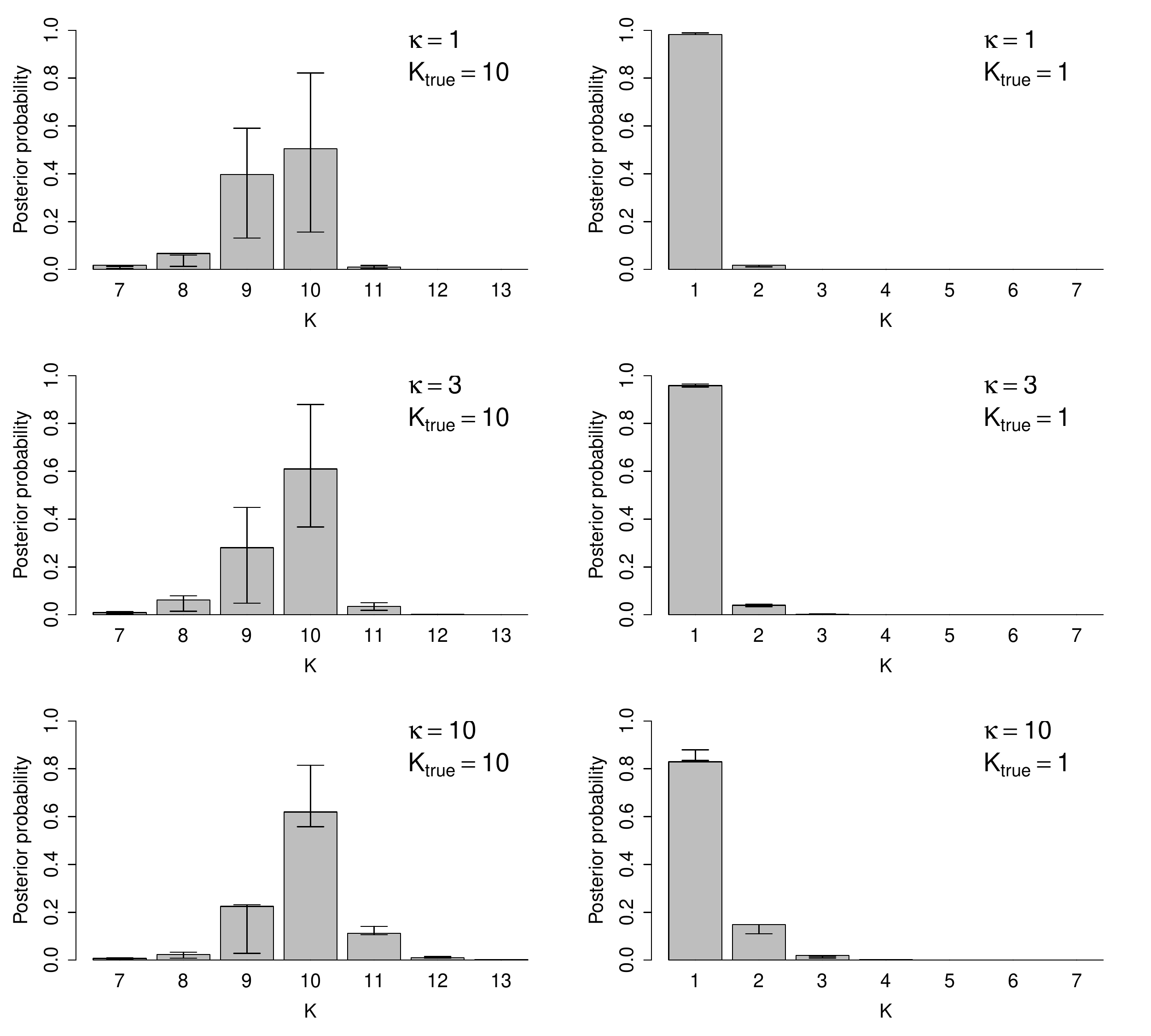}
\caption{Average posterior probabilities of plausible values of $\nk$ across ten datasets. Left plots show posteriors for the ten-source reality ($\nktrue=10$) with prior mean values of $\pk = 1, 3, 10$ from top to bottom. Right plots show posteriors for the one-source reality ($\nktrue=1$) with $\pk = 1, 3, 10$. In each plot, the 25\% and 75\% quantiles across the 10 datasets are indicated by the vertical error bars for each value of $\nk$.\label{fig:110source_test}}
\end{figure*}

For each of the 20 simulated datasets, ten RJMCMC chains were run {\coledits to assess convergence, but for simplicity only one chain per dataset was used in the final analysis.}\footnote{For the purposes of convergence diagnostics, we initialized each chain by randomly choosing between 1 and 20 sources and then deterministically spreading them out around the edge of the image space.} {\coledits The chains were run for 200,000 RJMCMC iterations, the first 100,000 of which formed the convergence period (or burnin) and were discarded.}  For each dataset, the posterior probability of being in state $\nk=\vk$ was calculated, using Equation \ref{eqn:mc_kest}, for all feasible values of $\vk$. Figure \ref{fig:110source_test} summarizes the inference for $\nk$ under the ten-source ($\nktrue=10$, left panels) and one-source ($\nktrue=1$, right panels) realities, for $\pk=1 ,3$ and $10$ (top, middle and bottom panels respectively). Recall that $\pk$ is the prior mean number of sources. The 25\% and 75\% quantiles of the posterior probabilities across the ten datasets are indicated for each value of $\nk$.

Figure \ref{fig:110source_test} shows that, for the ten-source reality, the posterior probability is concentrated around $\nk=9,10$ and $11$, regardless of which of the three values of $\pk$ is used. Indeed, the prior probability of ten sources specified by the prior with $\pk=10$ is nearly 1.25 million times that of the probability specified by the prior with $\pk=1$. Despite the difference in the prior probability as a function of $\pk$, the posterior probabilities of $\nk=10$ are quite consistent; the average (across simulations) differs by only about $0.1$ (comparing $\pk=1$ with $\pk=10$, see Figure \ref{fig:110source_test}). In other words, there is about a 1.25  multiplicative increase in the posterior probability of ten sources when the prior mean is changed from $\pk=1$ to $\pk=10$. This modest difference in posterior probability is acceptable as it is unlikely that prior information would allocate the truth 1.25 million to one odds.

There are appreciable differences among the simulated datasets as indicated by the quantiles in Figure \ref{fig:110source_test}. This is to be expected because the source positions and intensities are chosen randomly. Some of the simulated datasets have two sources very close to each other, making it hard to determine that they are distinct. In some cases, it is possible to separate these very close sources based on the spectral data (using the full model), i.e., if the spectral data appear to come from two {\sl gamma} distributions rather than one. However, in other cases it is difficult to separate such nearby sources, even with the spectral data. Indeed, checks confirmed that datasets with sizeable posterior probability at $\nk=9$ under the full model include overlapping sources that cannot be separated by eye and have similar spectral distributions. Posterior probability at $\nk$ values of 11 and above appear because chance clusters of photons are sometimes mistaken for separate sources. The precise location of these `ghost' sources, however, is highly erratic across RJMCMC iterations. There is limited evidence for them in the data and thus wide error bars for their ``locations" in the posterior distribution.

Inference is also robust to the choice of $\pk$ under the one-source reality ($\nktrue=1$). The posterior mode is clearly $\nk=1$ for all three values of $\pk$. Owing to the skewness of the Poisson density, the difference in prior probability of $\nk=1$ across the different $\pk$ values is less dramatic than that for $\nk=10$. When $\pk=1$ the {\it a priori} probability of $\nk=1$ is around 800 times that when $\pk=10$. Consequently, the difference in posterior probabilities is also less noticeable. Indeed, the qualitative difference in the posteriors under $\pk=1$ and $\pk=10$ is marginal, see Figure \ref{fig:110source_test}.

Our key conclusion is that the posterior probability of the true number of sources $\nk$ seems insensitive to the prior probability assigned to $\nk$, at least when using the Poisson prior. Consequently, the value of $\pk$ only needs to be in the region of the true number of sources in order for the fit for $\nk$ to be reasonable. These conclusions match our intuition that knowing the precise PSF statistically constrains the mixture model sufficiently for the data to drive the fitted values of the parameters. {\coledits Our simulations are representative of typical datasets, but establishing similar conclusions for smaller datsets may require more studies.} {\coledits A dataset could also be larger than those in our simulations,} but as $\pk$ (and $\nk$) increases, greater Poisson variance means that the absolute deviation of $\pk$ from the true number of sources has progressively less influence on posterior inferences. (Intuitively, it is more reasonable to {\it a priori} suspect 101 sources when there are 110, than to suspect 1 when there are 10). In our context, prior information typically consists of previous observations, possibly from a different wavelength band. Therefore, it can be assumed that the information is quite reliable and gross prior `misspecification' is unlikely. Clearly, priors other than the Poisson distribution can be considered if a more diffuse prior distribution is desired.

\subsection{Utility of the spectral model}
\label{sec:simulation}

Here we investigate the performance of our model and methods for a range of background intensities, source separations and relative source intensities. We compare the performance of the spatial-only and full models. For simplicity, we simulated data for a two-source ($\nktrue=2$) reality. In each simulation, the number of photons from the background and the number from each source were drawn from Poisson distributions with respective means $\sint_0,\sint_1,\sint_2$. We set $\sint_2 = 1000$ and $\sint_1 = \sint_2/\rint$, for $\rint=1,2,5,10,50$. We refer to $\rint$ as the relative intensity of the two sources. To set $m_0$ and quantify the strength of the simulated background in an astronomically meaningful way we define a source region in terms of the PSF. Specifically, we again use the King profile PSF and define the source region as the region with PSF greater than 10\% of its maximum. (The King profile density has no finite moments). We next define $\psr$ to be the probability that a photon from a source falls within its source region and set the background per source region to be $\sint_0 = \bint \psr \sint_2$, for $\bint= 0.001,0.01,0.1,1$. That is, the mean number of background photons in the faint source region was varied between $1/1000$ and $1$ times the mean number of photons from the faint source falling in the same region. As we shall discuss and unsurprisingly, the faint source was difficult to locate in datasets that were simulated with $\bint = 1$ and less so for those simulated with $\bint = 0.001$. Finally, the separation of the two sources was set to be $0.5,1,1.5$ or $2$ distance units. These separations can be interpreted using the fact that our source regions are approximately circles of radius 1.

\begin{figure*}[p]
\centering
\includegraphics[width=0.95\textwidth]{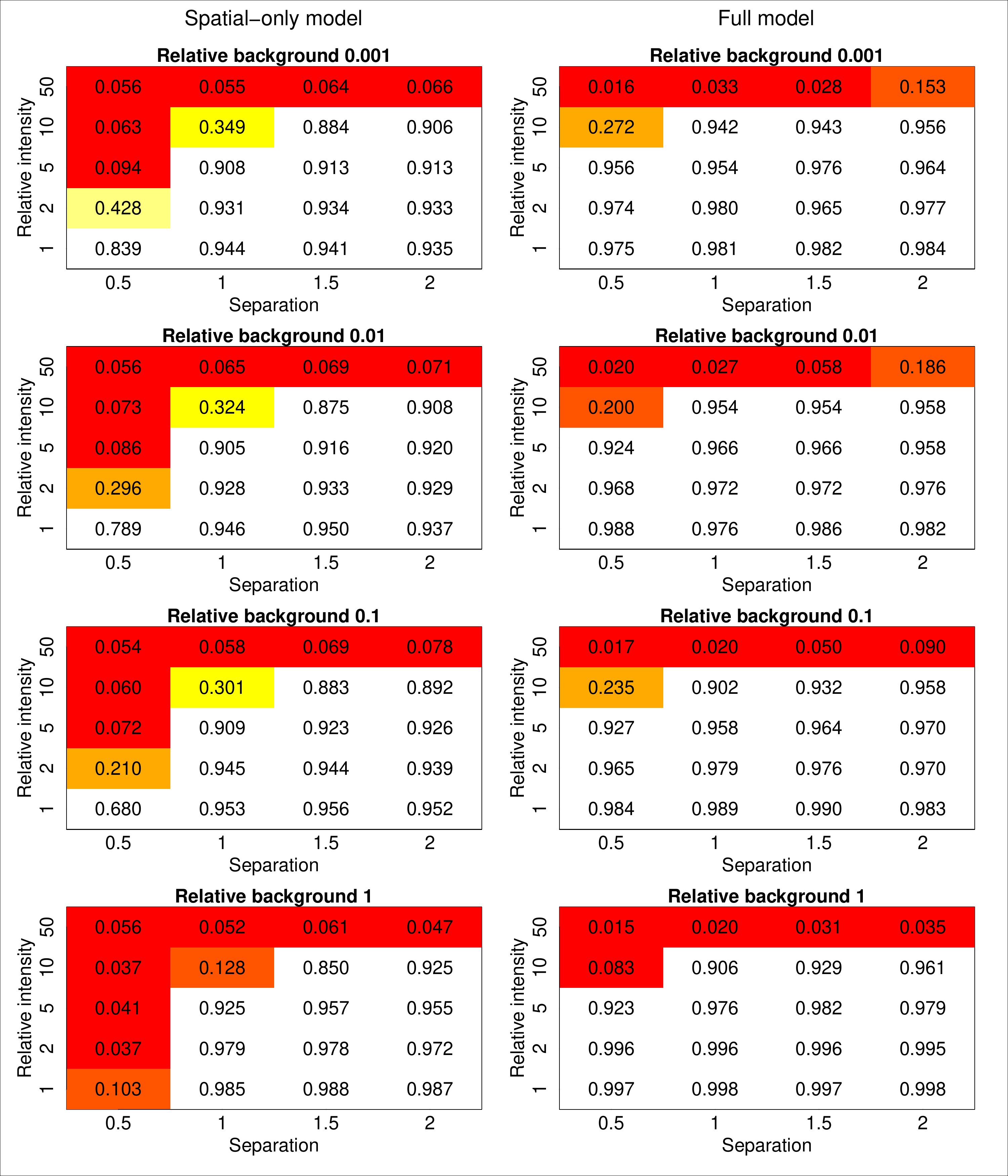}
\caption{Exploring the sensitivity of our algorithm to source separation, relative strengths, and background level. The median posterior probability of $\nk=2$ across the 100 simulations is shown; $\nktrue=2$ in all cases. The results from the spatial-only model (left column) and the full model (right column) are both shown. Red indicates probabilities less than $0.1$, and white indicates probabilities greater than $0.5$. (Intermediate colors indicate probabilities between $0.1$ and $0.5$.)  \label{fig:post2}}
\end{figure*}

Spectral data was also simulated for source and background photons. An aim of this simulation study aims is to investigate how much using the spectral data improves the fitted parameters. Since sources can only be distinguished by their spectra if their spectra are different, we used different spectra for the two simulated sources; specifically we set $\epa_1 = 3$, $\gm_1=600$, $\epa_2=6$ and $\gm_2=1500$.

In summary, our simulation study consists of a $5\times 4 \times 4$ grid of configuration settings ($\rint=1,2,3,5,10,50$; $\bint= 0.001,0.01,0.1,1$; and source separations of $0.5,1,1.5,2$). One hundred datasets were simulated for each of the resulting $80$ configurations, and analyzed using first the spatial-only model and then the full model. {\coledits In particular, for each dataset our algorithm was run for 20,000 RJMCMC iterations, the first 10,000 of which formed the convergence period (or burnin) and were discarded.\footnote{\coledits This is a relatively small number of RJMCMC iterations, but since our simulated datasets were quite small images each including only two sources, we found it to be sufficient.}} The median posterior probability of two sources is shown in Figure \ref{fig:post2} for each of the different simulation settings. The left and the right panels correspond to the spatial-only and full models, respectively. We use the median posterior probability across the 100 simulated datasets because in a few simulations the faint source is unusually bright or unusually faint, which noticeably effects the mean posterior probability of two sources. Nevertheless, summaries based on the mean posterior probability are qualitatively very similar, albeit with slightly more noise. We have organized the results by background intensity because in practical applications background is often well determined.

In images simulated with relative intensity $50$ the posterior probability of two sources tends to be low. This is because $\rint=50$ corresponds to a faint source intensity of $\sint_2=20$, while the brighter source has intensity $\sint_1=1000$. Thus, the faint source is typically not bright enough to be distinguished from noise; its photons can be adequately explained as a random cluster formed of photons from the brighter source or the background. In this case the posterior probability peaks sharply at $\nk=1$. The spatial-only model is more likely to mistake a cluster of background photons for {\it a} faint source and therefore, in the case of $\rint=50$ and small source separation, typically gives slightly higher posterior probabilities of two sources than the full model (but the probabilities are still very small). For less extreme relative intensities, using the full model increases the posterior probability of two sources. The improvement is particularly noticeable for relative intensities $5$ and $10$, regardless of the background strength. The spectral distribution of source counts reduces the plausibility that the faint source is just a cluster of photons from the background or the bright source. When both sources are bright and reasonably separated both the spatial-only and full models give high posterior probability at $\nk=2$.\footnote{One curiosity, present in the left panels of Figure \ref{fig:post2} (spatial-only model), is that when both of the sources are reasonably bright, greater median posterior probability of two sources is obtained when the background is {\it stronger}. This phenomenon occurs because, in the presence of strong background, deviations between the PSF and the observed counts are difficult to detect, whereas, with weak background, such deviations may be attributed to spurious additional sources. (Indeed, the posterior probability of $\nk=3$ is typically greater at low background levels than at high background levels). When the full model is used this effect is diminished. The curiosity is not qualitatively important because the bright sources are well identified in all cases.  Clearly weaker background is preferred as it improves the chance of detecting (real) faint sources.}

\begin{figure*}[p]
\centering
\includegraphics[width=0.85\textwidth]{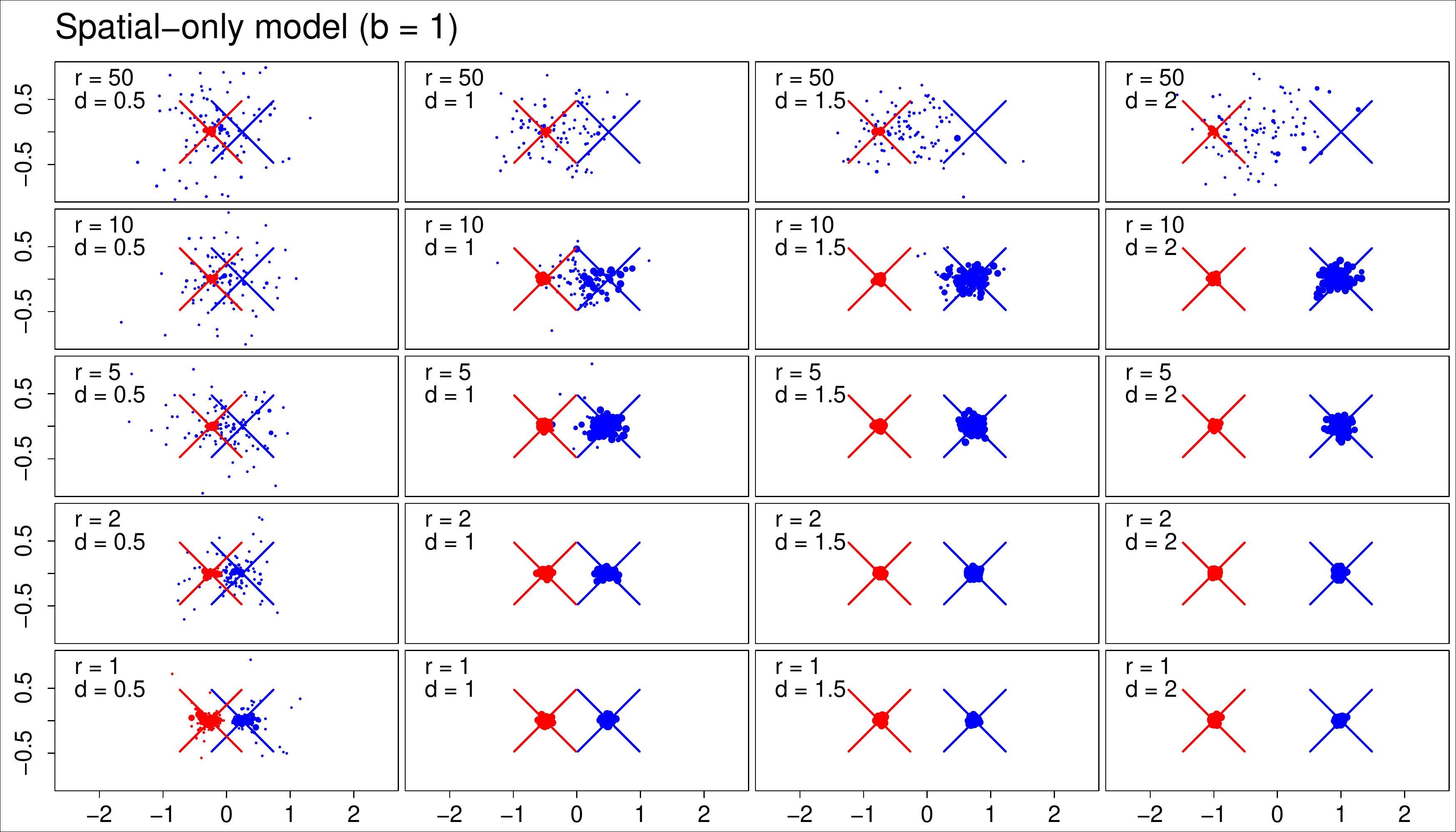}
\includegraphics[width=0.85\textwidth]{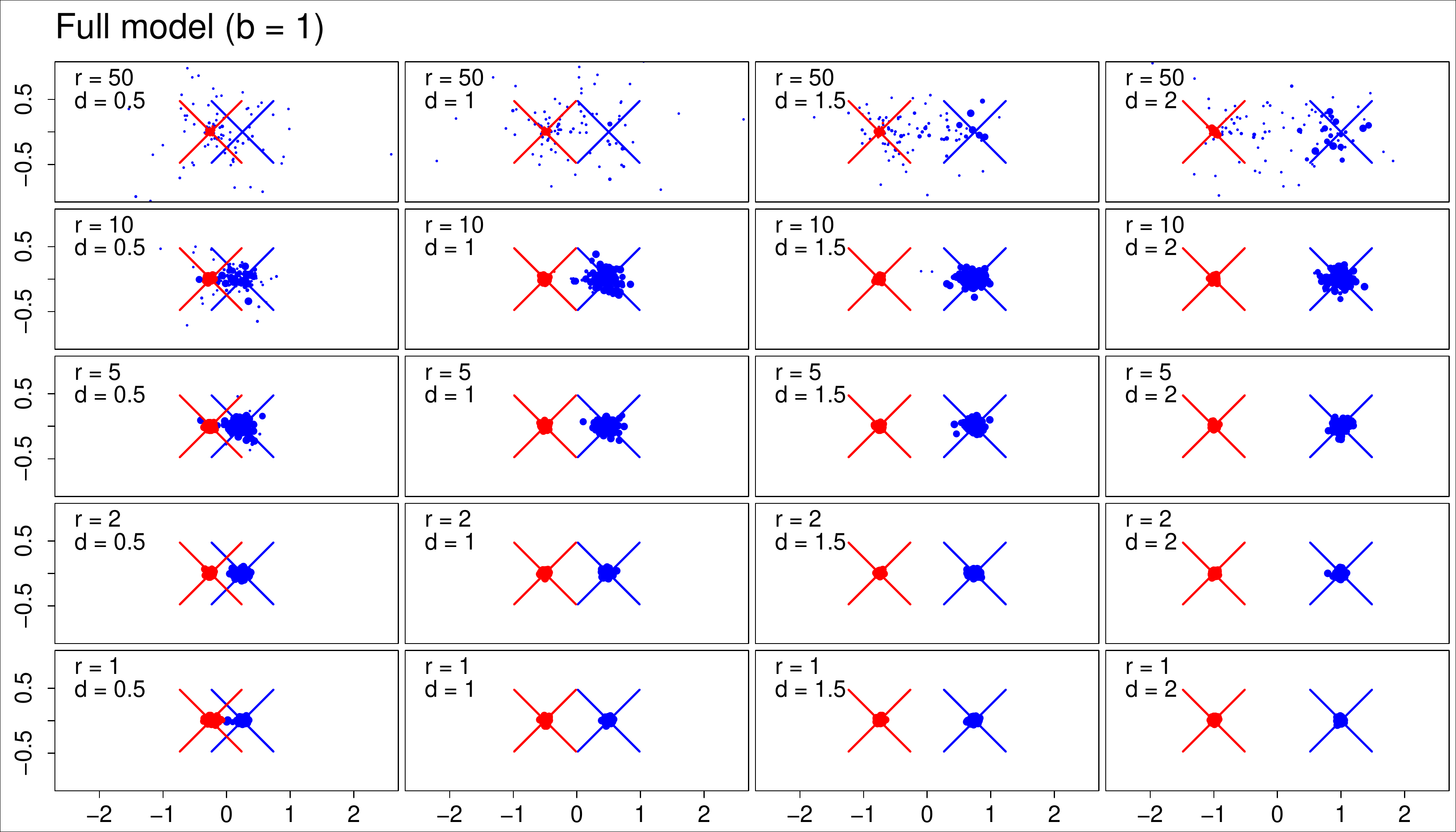}
\caption{Sensitivity of location determination as a function of source separation, relative strength, and
background level. The simulation is the same as that in Figure \ref{fig:post2}. Mean posterior locations of two sources for each of 100 simulations, under the spatial-only model (top 20 plots) and the full model (bottom 20 plots).  Red and blue dots give the mean posterior locations for each simulation of the bright and faint sources respectively. The large `X's of corresponding color indicate the true locations. The diameters of the dots are proportional to the posterior probabilities of two sources. The relative background, relative source intensity, and source separation are indicated by $b$, $r$ and $d$ respectively.\label{fig:pos}}
\end{figure*}

To fit the source parameters, we fix $\nk$ at its posterior mode value and use Equation \ref{eqn:generalk_estimate}. Although the fitted parameters of the bright source are always accurate, those for the faint source may be poor, especially if the posterior mode of $\nk$ is at 1 or if `ghost' sources have appreciable posterior probability. The accuracy of the faint source's fitted parameters essentially follows the pattern seen in Figure  \ref{fig:post2}. When the real faint source is very weak or located too close to the bright source, then a fitted second source (when the posterior mode of $\nk$ is greater than $1$) is likely to be a `ghost' consisting mainly of a cluster of photons from the background or the bright source. In which case, its fitted parameters bear little resemblance to those of the true faint source. This is illustrated in Figure \ref{fig:pos}, which shows the mean (conditional on $\nk=2$) posterior locations of the two sources for all 100 datasets under each configuration of simulation settings. Crosses indicate the true locations of the sources. The mean posterior locations of the bright source (red dots) are not always visible in the plots because they are often in the middle of the red crosses. The location of the bright source becomes slightly harder to fit as the intensity of the faint source increases. (This is at least partly because the background intensity is proportional to the faint source intensity). The size of the dots indicate the posterior probability of two sources.

The full model again yields more accurate fits. The fitted locations of the faint source (blue dots) center around its true location (blue crosses) for $\rint\leq10$, even when the source separation is small. For the spatial-only model there is more scatter. Under both models, when $\rint=50$ we can see that many of the fitted faint source locations correspond to spurious clusters of photons surrounding the bright source. As the separation increases some of the fitted faint source locations are halfway between the true locations of the two sources. This occurs when the posterior distribution of the faint source $x$-coordinate is bimodal, a spurious cluster of photons and the real faint source both being supported as possible second sources. In an actual analysis this bimodal behaviour would be apparent from inspection of the posterior draws of the source location. For $r=50$ and large separation, the full model sometimes accurately fits the faint source location, but the spatial-only model never does. The behavior of the other fitted parameters follows the same pattern illustrated in Figure \ref{fig:pos} because the fitted source locations indicate how well photons are allocated to the correct source. This is confirmed by inspecting tables of the mean (or median) squared error of each parameter (not shown).

The number of Monte Carlo samples used in estimating the mean posterior locations (conditional on $\nk=2$) is determined by the posterior probability of two sources, and thus is indicated by the size of the dots. Very small dots may have non-negligible Monte Carlo error i.e. the true posterior mean location (conditional on $\nk=2$) may be somewhat inaccurately approximated. This is because applying Equation \ref{eqn:generalk_estimate} for each parameter does not accurately compute the mean of $\post(\cparas_\nk|\nk,\datx,\daty,\en)$ for values of $\nk$ that have low posterior probability.\footnote{We could instead fix $\nk=2$ and run a standard MCMC algorithm to obtain a large enough posterior sample to accurately fit the mean posterior locations. We do not pursue this strategy because the fitted parameters that are conditional on unlikely values of $\nk$ are of little practical use.} However, in practice, when the number of sources is unknown, it makes sense to only consider values of $\nk$ with relatively high posterior probability. Furthermore, one typically checks the level of Monte Carlo error for the values of $\nk$ of interest, by running multiple chains. Large variation in the parameter estimates across the chains indicates high Monte Carlo error. In which case, one should run the chains longer in order to obtain a larger Monte Carlo sample.

\section{Application I: XMM dataset}
\label{sec:xmm}

\begin{figure*}[p]
\centering
\includegraphics[trim = 0mm 10mm 0mm 10mm,width=1\textwidth]{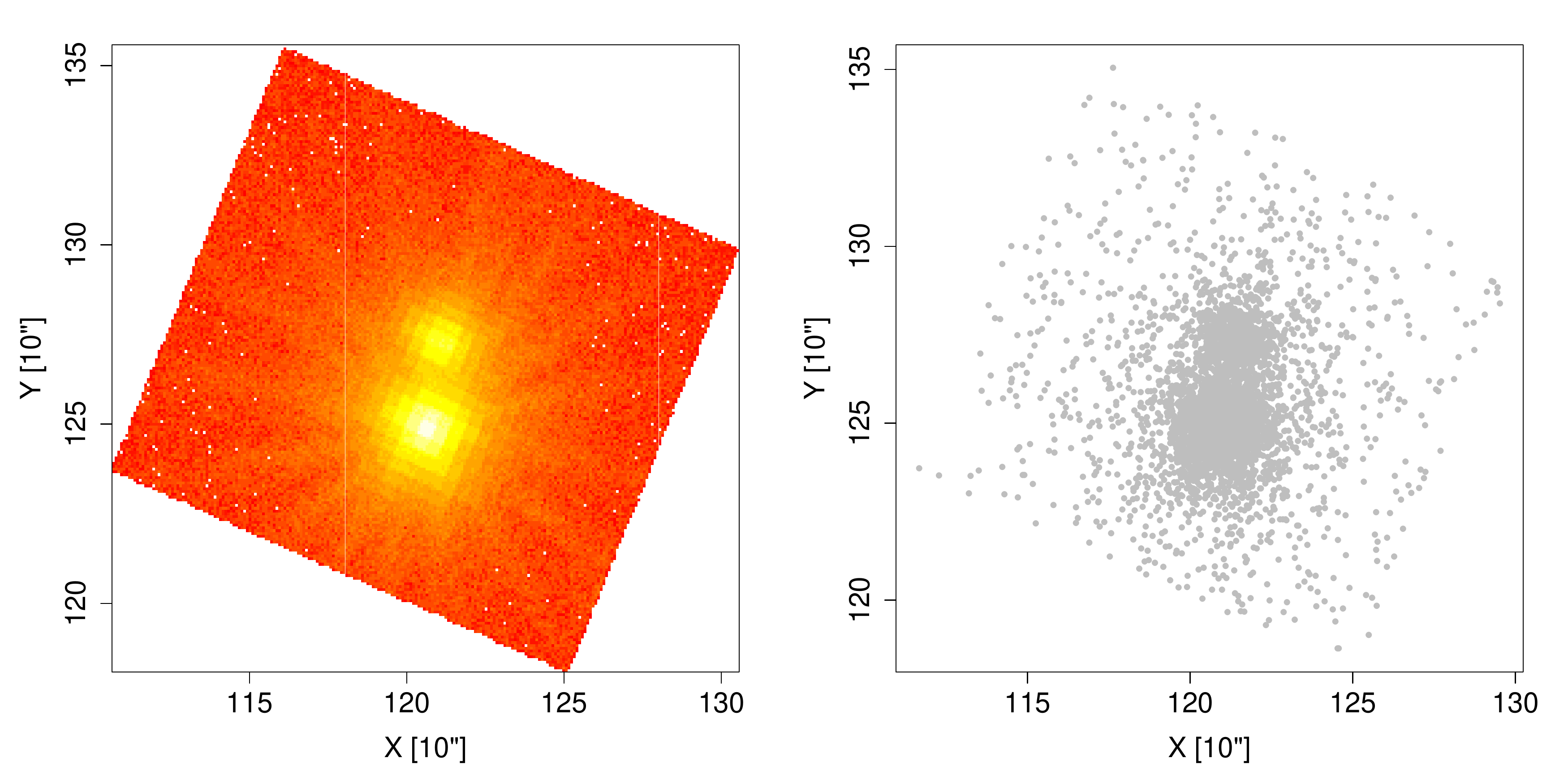}
\caption{Visual binary FK and FL Aqr observed with XMM-{\sl Newton} (FK is the brighter source at
bottom). The XMM obs\_id is  0151450101. Shown is a counts image with $10''$ bins and arbitrary origin (left), and a scatter plot of a
subset of 6,000 events over a 5ks subexposure (right).\label{fig:xmm1}}
\end{figure*}

\begin{figure*}[p]
\begin{center}
\includegraphics[trim = 0mm 10mm 0mm 5mm,width=0.6\textwidth]{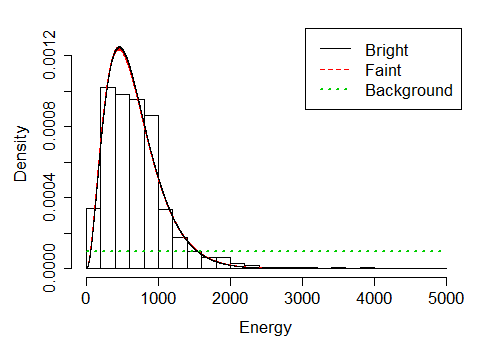}
\end{center}
\caption{A histogram of the spectral data in the XMM observation of FK Aqr and FL Aqr. Plotted are 1,000 spectra for the bright (solid black lines) and faint (dashed red lines) sources, each corresponds to a posterior sample of the spectral parameters. (The posterior variance is small on this scale.) The background spectra is shown by the dotted green line.\label{fig:xmm_spectral}}
\end{figure*}

We now apply the spatial-only and full models to an XMM observation (obs\_id  0151450101) of the apparent visual binary FK Aqr and FL Aqr. The data consist of the spatial and spectral information of around 540,000 photons detected during a 47ks exposure. The spatial data is displayed in Figure \ref{fig:xmm1} as both an image (left) and a scatter plot (right), and the spectrum is plotted in Figure \ref{fig:xmm_spectral}.  The moderate overlap of the sources and high counts make this a good test of our model. In particular, we expect that the spatial-only model and full model analyses to be similar (for the spatial parameters) because of the large amount of spatial information. Furthermore, since the data clearly indicate two sources, we can concentrate on verifying that our model yields sensible posterior inference using standard MCMC. (This gives draws from the joint posterior for a fixed number of sources and therefore results in inference that is simpler to interpret than inference resulting from RJMCMC.) Use of the more complicated RJMCMC analysis is reserved for the {\sl Chandra} dataset in Section \ref{sec:chandra} because there is non-negligible uncertainty in $\nk$ for that dataset.

\begin{table*}[t]
 \caption{Posterior means under the spatial-only model and the full model.
 The parenthetic intervals are $1\sigma$ error bars computed using 16\% and 84\% posterior quantiles.\label{tab:parasxmm}\vspace{0.25cm}}
\centering
\resizebox{0.6\linewidth}{!}{
\begin{tabular}{l|rl|rl}
  \hline
 & \multicolumn{2}{c|}{Spatial-only model} & \multicolumn{2}{c}{Full model} \\
\hline
$\mu_{1x}$  & 120.974 & (120.973,120.975) & 120.973 & (120.973,120.974) \\
$\mu_{1y}$   & 124.873 & (124.873,124.874) & 124.873 & (124.872,124.874) \\
$\mu_{2x}$ & 121.396 & (121.394,121.398) & 121.397 & (121.395,121.399) \\
$\mu_{2y}$ & 127.319 & (127.317,127.321) & 127.326 & (127.324,127.328) \\
 $w_1$   & 0.717 & (0.716,0.718) & 0.732 & (0.731,0.732) \\
$w_2$ & 0.182 & (0.181,0.182) & 0.189 & (0.189,0.190) \\
$w_0$ & 0.102 & (0.101,0.102) & 0.079 & (0.079,0.079) \\
 $\gamma_1$  & \multicolumn{1}{r}{--} & \multicolumn{1}{l|}{--} & 664.86 & (664.43,665.30) \\
$\gamma_2$ &\multicolumn{1}{r}{--} & \multicolumn{1}{l|}{--} & 662.78 & (661.78,663.87) \\
$\alpha_1$  & \multicolumn{1}{r}{--} & \multicolumn{1}{l|}{--} & 3.205 & (3.199,3.211) \\
$\alpha_2$ &\multicolumn{1}{r}{--} & \multicolumn{1}{l|}{--} & 3.131 & (3.118,3.144) \\
   \hline
\end{tabular}}
\end{table*}

In the image shown on the left of Figure \ref{fig:xmm1} the sources seem to have faint `spokes'. Approaches for modeling these features are suggested in \citet{read2010} and  \citet{read2012}, but we use the unaltered King profile PSF for simplicity.  As mentioned in Section \ref{sec:data}, the spatial data are binned when recorded on the observatory LCD screen. However, the bins are small in comparison to the XMM PSF so our use of a model that treats the data as unbinned is reasonable. (See Section \ref{sec:summary} for further discussion.)

For the spatial-only model and the full model, ten MCMC chains (with $\nk$ fixed at 2) were run  {\coledits for 20,000 MCMC iterations, the first 10,000 of which formed the convergence period (or burnin) and were discarded.\footnote{{\coledits Note that, since we used standard MCMC and there are only two bright sources, the number of MCMC iterations until convergence was relatively small.}}} The large amount of data means that the source locations are precisely fit by both models, as can be seen in Table \ref{tab:parasxmm}. However, the posterior mean of the relative intensity of the background is about 20\% lower for the full model. This is presumably due to a greater ability to separate source and background counts with the additional information given by the spectral data. In particular, photons from the sources can be found across the entire image so there is a tendency to over-estimate the background intensity without some non-spatial way of distinguishing its photons from those of the sources.

Until now, it has not  been possible to distinguish the spectral distributions of these two sources. Conventional fitting of the spectra extracted from non-overlapping source regions give statistically indistinguishable results, with identical column density $N_{\rm H}{\approx}1.0-1.6$ $(10^{20}$~cm$^{-2})$, double temperature components $kT_1 {\approx} 0.25-0.26$~(keV), $kT_2 {\approx} 0.78-0.82$~(keV), and metallicities $Z{\approx} 0.12-0.14$. This remarkable coincidence could be attributed to strong contamination of FL Aqr by photons from FK Aqr. Our algorithm, which eliminates such contamination, can answer the question of how similar the two sources are. Of course, a comparison of the source spectra shapes is only possible using the full model. Figure \ref{fig:xmm_spectral} shows 1,000 spectra sampled from the posterior distribution\footnote{To reduce correlation, every $10^{\rm th}$ sample of the original 10,000 {\coledits stored} MCMC samples of the spectral parameters was used.} for the bright (black solid lines) and faint (red dashed lines) sources; for each source, all 1,000 spectra are very similar and so appear as a single curve. We observe that the bright source spectra are very similar to the faint source spectra, which is consistent with the difficulty  in distinguishing the spectral distributions of the two sources in previous analyses.

\begin{figure*}[t]
\begin{center}
\includegraphics[trim = 0mm 10mm 0mm 15mm,width=1\textwidth]{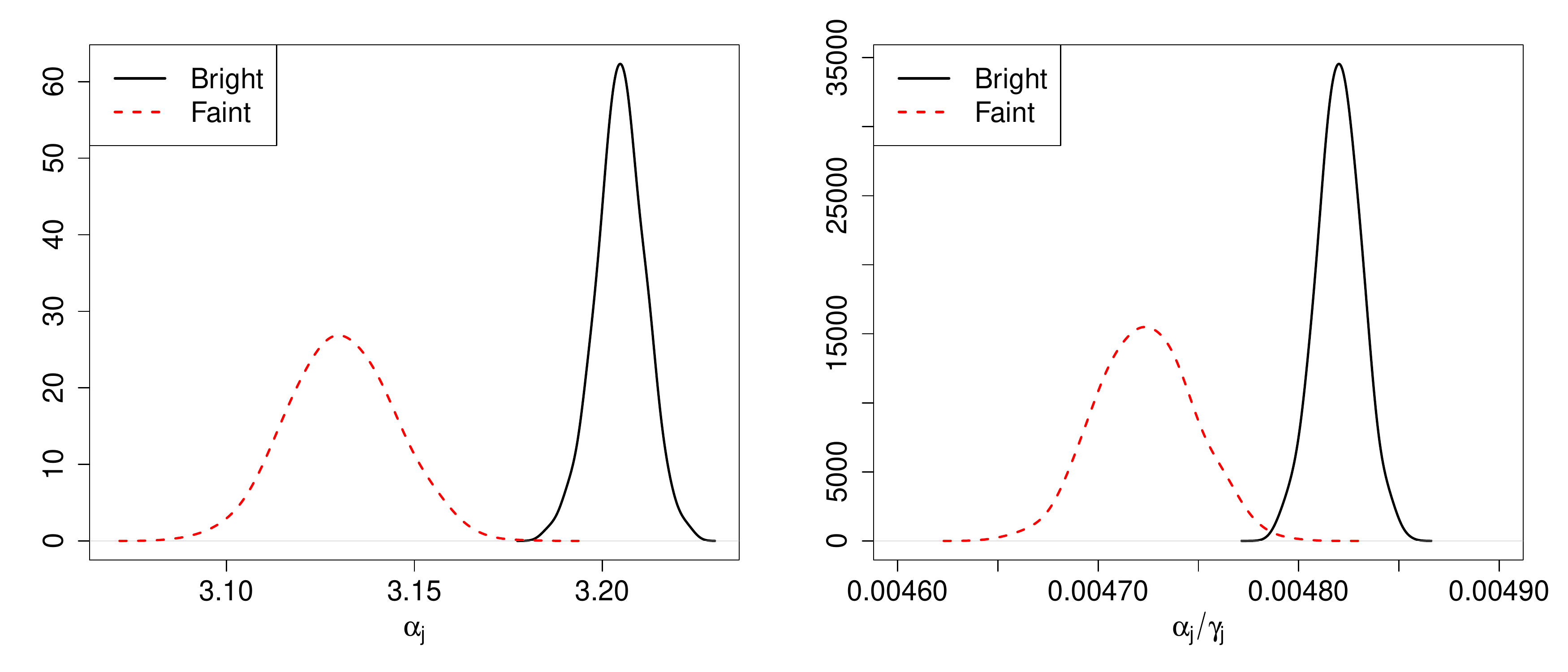}
 \caption{Posterior distributions of the parameters of the {\sl gamma} distributions used to model the spectra of FK Aqr and FL Aqr. {\coledits The posterior distributions of the shape and rate parameters are shown in the left and right panels, respectively.}  \label{fig:xmm_spectral_paras}}
   \end{center}
\end{figure*}

Although the overall shapes of the two spectral are similar (Figure \ref{fig:xmm_spectral}), we can distinguish them by examining the parameters of their underlying {\sl gamma} distribution.  Figure \ref{fig:xmm_spectral_paras} plots the posterior distributions of these parameters for the two sources and shows that they clearly differ. We have plotted the shape and rate parameters, because the shape and variance differ more than the mean. The posterior distributions in Figure \ref{fig:xmm_spectral_paras} indicate that there is very little uncertainty in the spectral parameters; the intervals in Table \ref{tab:parasxmm} convey a similar message. This precision is obtained because of the large amount of data combined with the fact that our method properly accounts for uncertainty in photon origins and jointly fits spectral and spatial parameters.  Although our analysis is only physically accurate to the extent that the source spectra can reasonably be modeled with {\sl gamma} distributions, it nevertheless provides evidence that the spectra do differ in some way. More detailed conclusions would be possible with a physics-based spectral model that accounts for emission lines and other spectral features. A possible extension of this work is to replace the {\sl gamma} spectral model with a more complete model. A computationally less intensive approach is described in Section \ref{sec:secondary_analysis}.

\section{Application II: Chandra dataset}
\label{sec:chandra}

We analyze a {\sl Chandra} observation of the Orion Nebula Cluster using the spatial-only model and the extended full model given in Equation \ref{eqn:likelihood_extended}. The extended full model is used because the full model is not sufficiently flexible to capture the shape of the source spectra, as explained in Section \ref{sec:spectral_extensions}. The specific dataset we analyze is a subset of ObsID 1522 that omits the central source, a region where the PSF is distorted due to strong pile-up (Figure \ref{fig:chandra_positions}). The data include events that occurred within the first 20ks of the observation, of which there are $\approx14,000$.

\subsection{Anaylsis using the spatial-only and extended full models}
\label{sec:chandra_analysis}

\begin{figure*}[p]
\begin{center}
\includegraphics[trim = 0mm 5mm 0mm 15mm, width=1\textwidth]{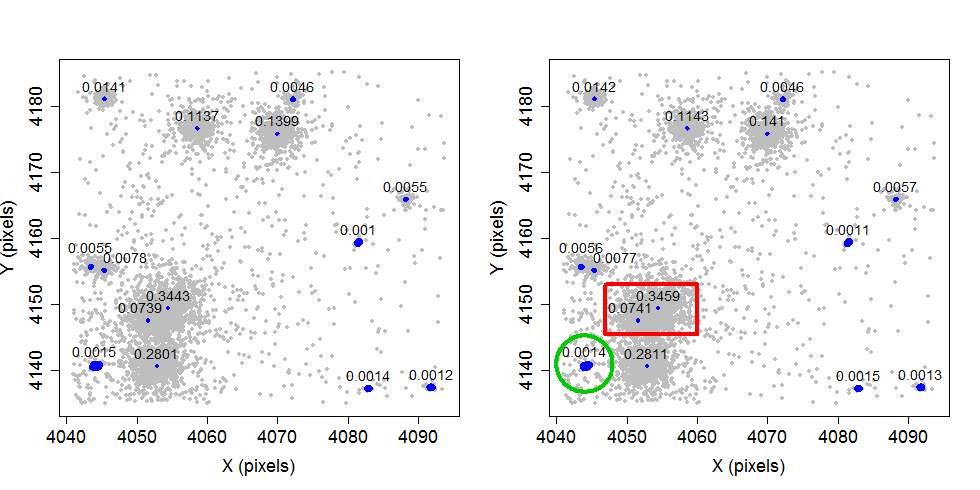}
\caption{Chandra observation of a crowded field near the center of the Orion Nebula Cluster. This field is approximately $25''{\times}25''$ in size, and is centered at (RA,Dec)=(5:35:15.4,-05:23:04.68). Shown in blue are approximate 90\% posterior credible regions for source locations, under the spatial-only model (left), and the extended full model (right). The figures next to the regions indicate the estimated relative intensities. The credible region of the source with the largest location uncertainty is circled in green (right panel). The red rectangular box encloses two overlapping sources (right panel) for
which we carry out a detailed follow-up spectral analysis (Section \ref{sec:secondary_analysis}). \label{fig:chandra_positions}}
  \end{center}
\end{figure*}

\begin{figure*}[p]
\centering
\includegraphics[trim = 0mm 10mm 0mm 5mm,width=1\textwidth]{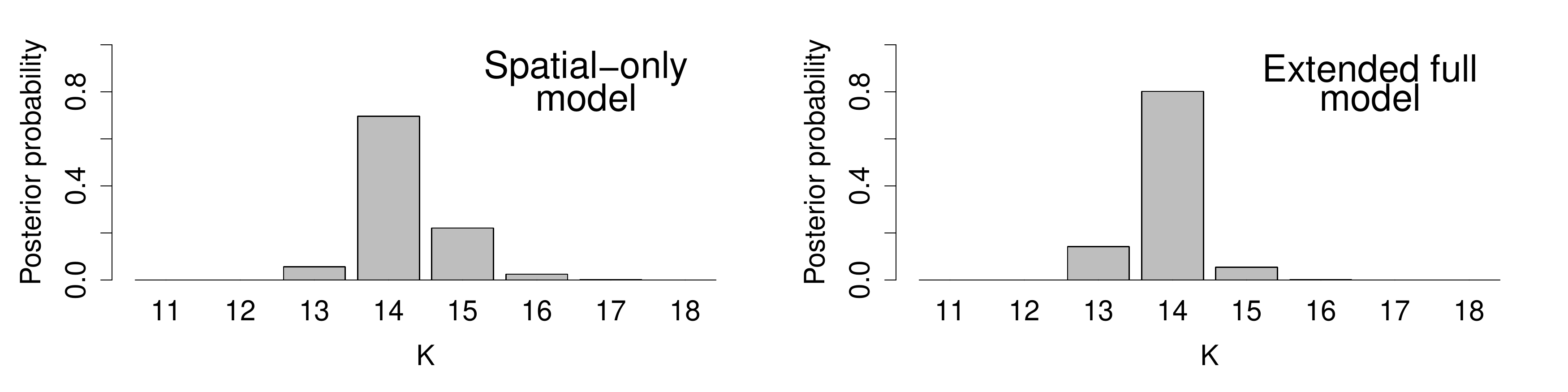}
\caption{
Number of sources detected in the analysis of the Chandra observation in Figure \ref{fig:chandra_positions}. Posterior of $K$ based on the spatial-only model (left) and the extended full model (right). \label{fig:chandrak}}
\end{figure*}

For both models, ten RJMCMC chains were run {\coledits for 150,000 RJMCMC iterations, the first 100,000 of which formed the convergence period (or burnin) and were discarded.} The posterior distribution of the number of sources, under the spatial and  extended full models, is displayed in Figure \ref{fig:chandrak}. The mode of both posteriors is at 14. However, the spatial-only model shows slightly more uncertainty, and some support for 15 sources.

As mentioned in Section \ref{sec:bayesian_analysis}, $\nk$  determines the number and meaning of the other model parameters and therefore we must condition on a value of $\nk$ to draw meaningful inferences for them from the RJMCMC output. Figure \ref{fig:chandra_positions} shows 90\% posterior credible regions (blue) for the locations of the sources under the two models, given $\nk=14$. Each credible region shows an area which has 0.9 posterior probability (given $\nk=14$) of containing the location of the relevant source, i.e., an integral of the posterior distribution of the source location (given $\nk=14$) over this area would evaluate to 0.9. The credible regions look to be similar under the two models. The estimated relative intensities also appear in Figure \ref{fig:chandra_positions} and are also similar, but are slightly lower under the spatial-only model for most sources. This is due to a higher estimate of the relative background intensity under the spatial-only model ($0.0053$ versus $0.0006$ under the extended full model\footnote{The background is likely inaccurately estimated by both models because the King profile PSF that we use is an approximation to the {\sl Chandra} PSF; the latter is more concentrated at its center. Thus, in our analysis, too many photons are allocated to the wings of the sources, deflating the background. That our analysis has still found genuine sources illustrates that it is not {\coledits too} sensitive to the PSF, {\coledits at least in the case of specifying overly heavy wings. If instead the raytraced PSF (ChaRT: {\tt http://cxc.cfa.harvard.edu/chart/}) is used, then the estimate of the background is higher because this PSF has lighter wings than the King profile. The lighter wings also lead to the detection of four additional faint sources: one has an optical counterpart, one does not, and two cannot be confirmed optically because they are close to a bright source. Further investigation of these sources and modeling possible variations in the PSF are topics for future work. For example, temporal information can potentially be used as a diagnostic to assess whether any of the detected weak sources are in fact due to fluctuations in the PSFs of the bright sources.}}).   Table \ref{tab:chandra_fits} gives the the posterior mean fit of the source locations and relative intensities under the extended full model for $\nk=14$. The detected sources are also matched to the source catalog from the Chandra Orion Ultradeep Project (COUP; \citealt{getman2005}).

\begin{table*}[t]
\caption{Extended full model fit for the Chandra observation in Figure \ref{fig:chandra_positions}. Posterior mean locations and relative intensities (as percentages), with 68\% intervals indicated. \label{tab:chandra_fits}\vspace{0.25cm}}
\centering
\resizebox{1\linewidth}{!}{
\begin{tabular}{c|rl|rl|rl}
  \hline
COUP \# & \multicolumn{1}{c}{$\p_{jx}$}& & \multicolumn{1}{c}{$\p_{jy}$}  & & \multicolumn{2}{l}{Relative intensity (\%)}\\
  \hline
732 & 4054.42 & (4054.41,4054.43) & 4149.45 & (4149.44,4149.46) & 34.59 & (34.16,35.03) \\
745 & 4052.83 & (4052.81,4052.84) & 4140.67 & (4140.66,4140.68) & 28.11 & (27.71,28.51) \\
689 & 4069.93 & (4069.91,4069.94) & 4175.93 & (4175.91,4175.94) & 14.10 & (13.79,14.40) \\
724 & 4058.57 & (4058.56,4058.59) & 4176.73 & (4176.71,4176.74) & 11.43 & (11.16,11.71) \\
744 & 4051.53 & (4051.50,4051.55) & 4147.57 & (4147.55,4147.60) & 7.41 & (7.14,7.68) \\
765 & 4045.40 & (4045.35,4045.46) & 4181.20 & (4181.15,4181.25) & 1.42 & (1.32,1.53) \\
649 & 4088.16 & (4088.08,4088.24) & 4165.95 & (4165.87,4166.03) & 0.57 & (0.50,0.63) \\
766 & 4045.36 & (4045.27,4045.45) & 4155.18 & (4155.10,4155.25) & 0.77 & (0.68,0.87) \\
788 & 4043.48 & (4043.36,4043.61) & 4155.74 & (4155.64,4155.84) & 0.56 & (0.47,0.64) \\
682 & 4072.11 & (4072.01,4072.21) & 4181.12 & (4181.03,4181.22) & 0.46 & (0.39,0.52) \\
640 & 4091.73 & (4091.53,4091.92) & 4137.42 & (4137.26,4137.59) & 0.13 & (0.10,0.16) \\
664 & 4081.43 & (4081.22,4081.63) & 4159.41 & (4159.21,4159.61) & 0.11 & (0.08,0.14) \\
665 & 4082.84 & (4082.67,4083.02) & 4137.28 & (4137.14,4137.43) & 0.15 & (0.12,0.19) \\
779 & 4044.39 & (4043.86,4044.60) & 4140.72 & (4140.43,4140.90) & 0.14 & (0.09,0.18) \\
Background & \multicolumn{1}{c}{--} & \multicolumn{1}{c|}{--} & \multicolumn{1}{c}{--} & \multicolumn{1}{c|}{--} & 0.06 & (0.01,0.10) \\
   \hline
\end{tabular}}
\end{table*}

Other observations of Orion suggest that the source circled (in green) in the right panel of Figure \ref{fig:chandra_positions} is a genuine source. Its location is more uncertain than other sources because it is more difficult to detect. Indeed, with an estimated intensity between 13 and 25 counts, this source is at the edge of {\coledits detectability} of local detection methods, particularly since the estimate of the local background in such methods would be high due to contamination from nearby bright sources. Thus, we expect that more basic approaches would either have failed to find this source, or would only find it by rendering their detection threshold to a point where spurious detections became problematic. Indeed, the reason the spatial-only model gives non-negligible weight to 15 sources (see Figure \ref{fig:chandrak}) is that it tends to split sources into two. The problem is that a single empirical PSF may exhibit chance variations that appear to be evidence for multiple PSFs. {\coledits The spatial-only model also mistakes clusters of background photons for sources. The locations and spectra of these spurious sources show considerable posterior variability. Although any particular instance has low probability, there are multiple instances that together create erroneous support for an additional source.} The main advantage of using the spectral information, in this example, is that it mitigates these issues, leading to a greater certainty that there are really 14 sources. Additionally, under the extended full model, the standard deviations of the parameters are almost invariably slightly smaller.

\subsection{Spectral analysis of the disentangled sources}
\label{sec:secondary_analysis}

The extended full model only captures the basic shape of the source spectra and we now illustrate how detailed follow-up spectral analysis can incorporate probabilistic event allocations. We perform this analysis for the two overlapping sources that are enclosed in the red box in the right panel of Figure \ref{fig:chandra_positions} (COUP sources \#732 and \#744 (\citealt{getman2005}). Their estimated relative intensities are $0.3459$ and $0.0741$ under the extended full model. This is a good example to test the probabilistic event allocations, since the sources are close together (separation ${\approx}1.7''$), each have sufficient counts for a useful spectral fit (${\approx}4350$ and ${\approx}910$ counts between $0.5-7$\,keV for the bright and faint sources, respectively), and one source is substantially weaker than the other.

As described in Section \ref{sec:more_model}, $\alloc_i$ indicates the source (or background) number associated with photon $i$. These are unknown parameters (or latent variables) that are updated at each iteration of the RJMCMC sampler. The variability in $\alloc_i$ indicates the uncertainty in the source of photon $i$ (due to the PSF and uncertainty in the source parameters). We can account for this uncertainty by conducting many spectral analyses, each according to a sampled photon allocation (i.e., sampled values of $\alloc_i$), and combining the results. We focus on photons with spatial location in the red box in Figure \ref{fig:chandra_positions} (right panel) and to values of $\alloc_i$ sampled conditional on $\nk=14$. Since we are only interested in COUP sources \#732 and \#744 we ignore any photons that are attributed to one of the other sources (in a given allocation). (The photons in the red box in Figure \ref{fig:chandra_positions} are attributed to one of the other sources only rarely).

\begin{figure*}[p]
        \centering
        \subfigure{\label{fig:a}\includegraphics[trim = 40mm 10mm 30mm 50mm,width=0.49\textwidth]{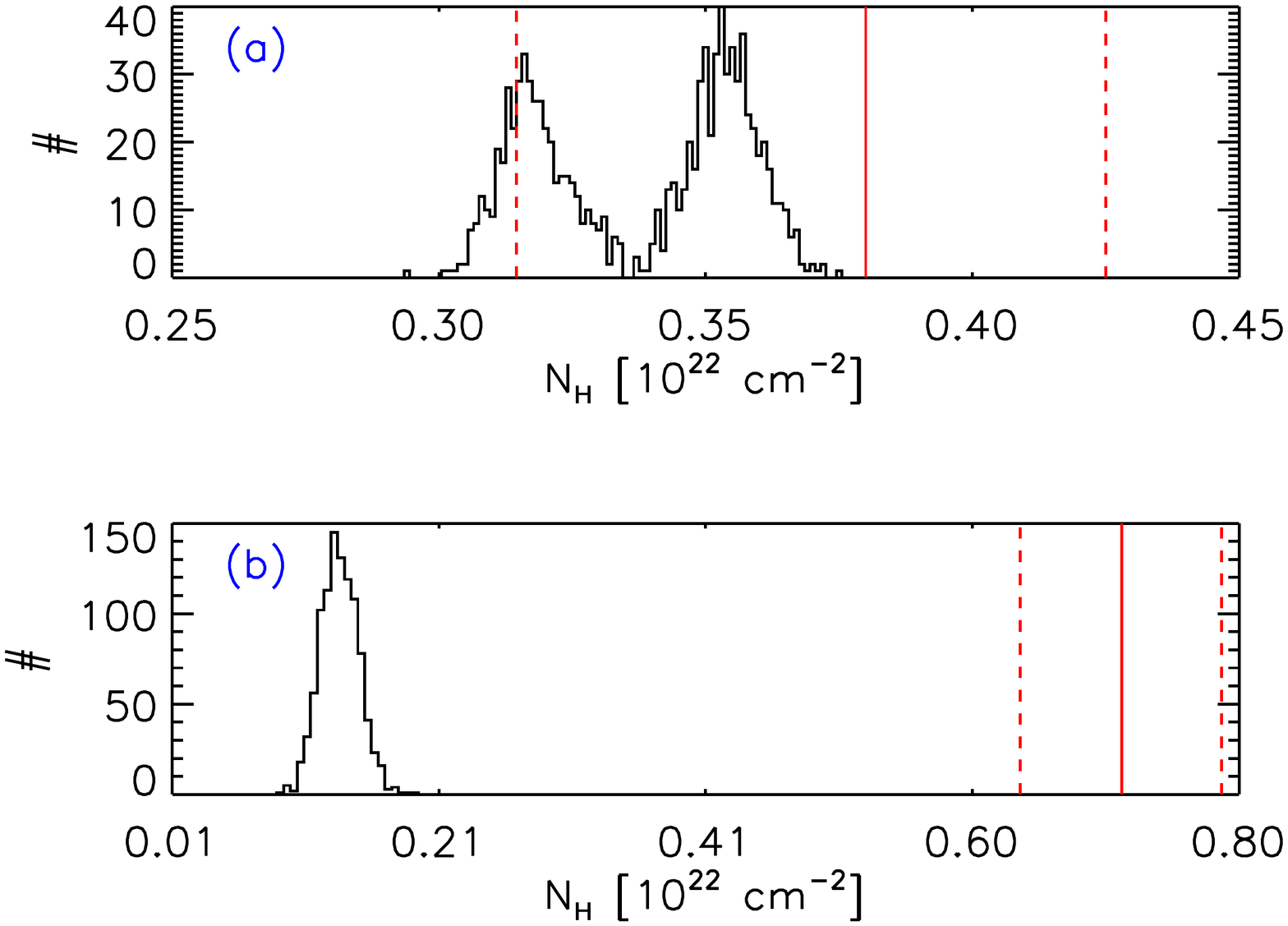}}
\subfigure{\label{fig:b}\includegraphics[trim = 40mm 10mm 30mm 50mm,width=0.49\textwidth]{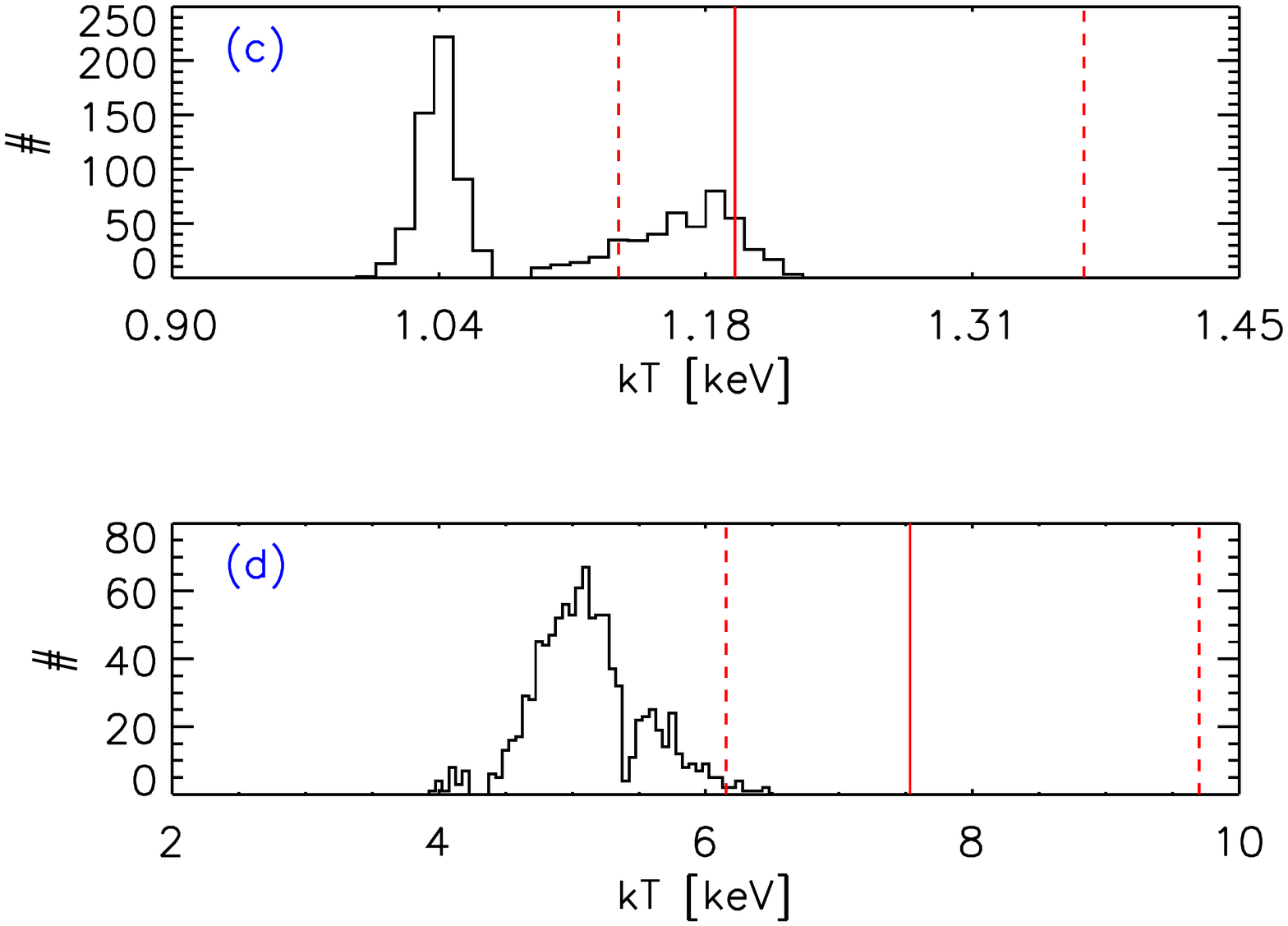}}\qquad
        \subfigure{\label{fig:a}\includegraphics[trim = 40mm 20mm 30mm 35mm,width=0.49\textwidth]{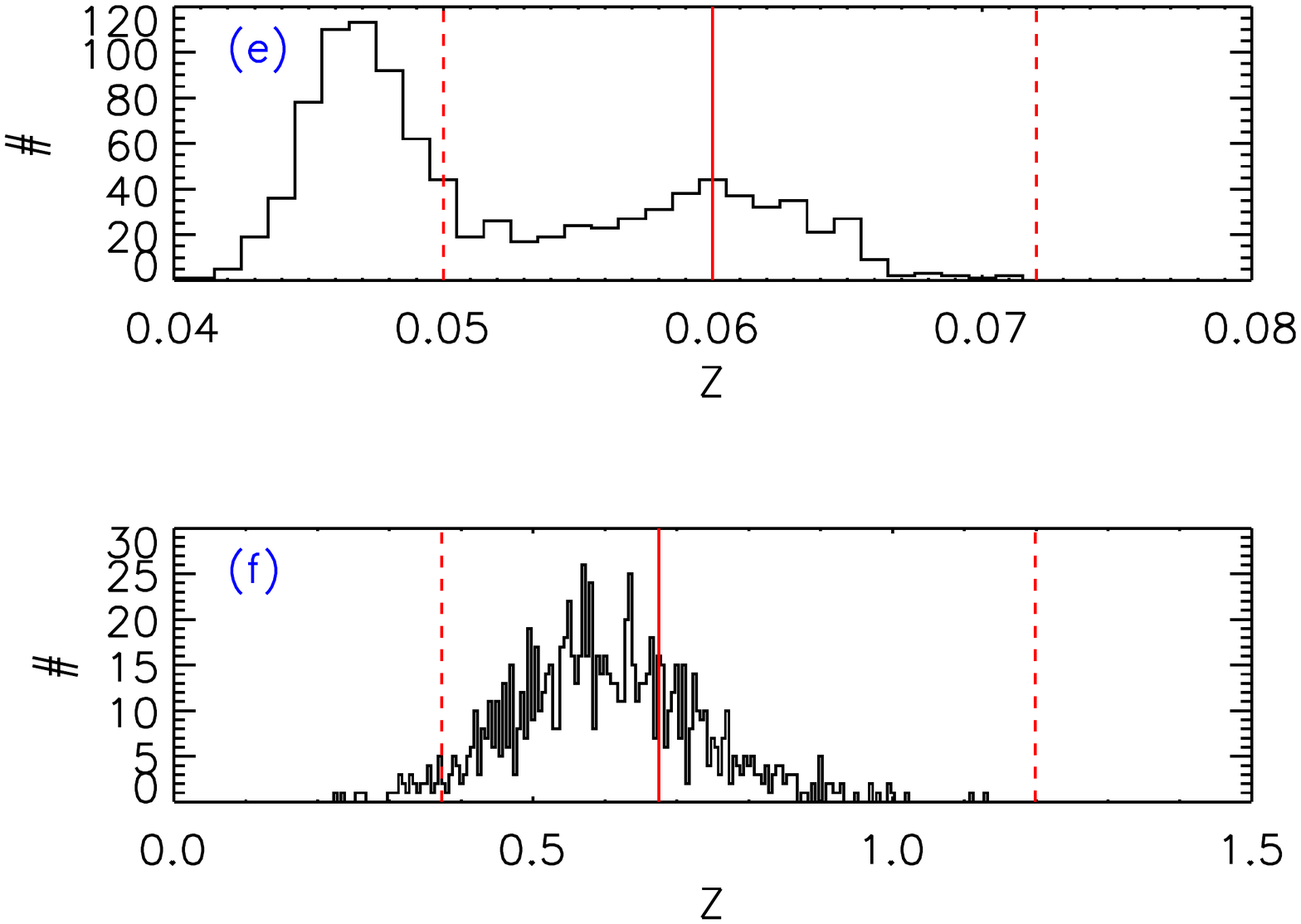}}
\subfigure{\label{fig:b}\includegraphics[trim = 40mm 20mm 30mm 35mm,width=0.49\textwidth]{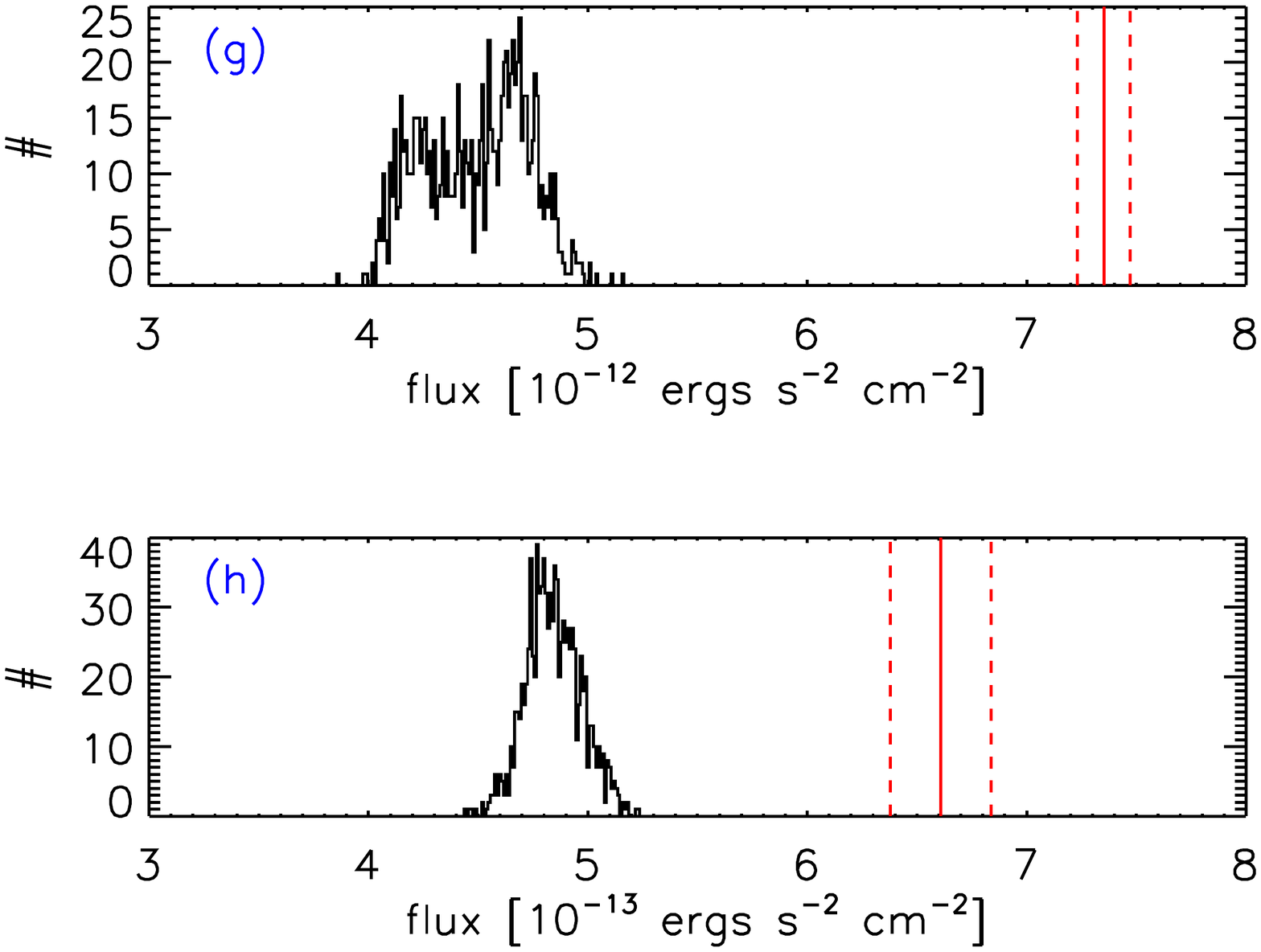}}
        \caption{Detailed spectral analysis of overlapping COUP sources \#732 and \#744. Best-fit values of absorption column ((a), (b)), temperature ((c), (d)), metallicity ((e), f)), and flux ((g), (h)) for the disentangled analysis, for each of 1000 allocations of the photons are shown as histograms. Panels (a), (c), (e), and (g) correspond to the bright source and panels (b), (d), (f), and (h) correspond to the fainter source.  The \naive\ analysis best-fit values and their 68\% intervals are shown by the  solid and dashed red vertical lines, respectively. The width of the histograms only account for uncertainty due to the allocation of photons, and not additional statistical error, which is well described by the intervals shown for the \naive\ analysis.\label{fig:secondart_analysis}}
\end{figure*}

Based on the photon allocations, we construct a sample of 1000 simulated spectral datasets for both sources, constructed from photon allocations based on every $10^{\rm th}$ iteration of the RJMCMC algorithm that sets $\nk=14$ (up to the $10,000^{\rm th}$ RJMCMC iteration that sets $\nk =14$). The variability in the source counts across the 1,000 iterations is ${\pm}17$ for both the bright and faint sources. The specific photons that are allocated to each source also varies, even when the total source counts do not. Each individual spectrum is fit with an absorbed single temperature thermal model ({\tt xsphabs*xsapec} in CIAO/{\sl Sherpa} v4.6) fitting the absorption column ($N_{\rm H}$), temperature ($kT$), metallicity ($Z$), and normalization. A pile-up correction is needed for all spectra for the bright source since the measured count rate of $0.7$~counts~frame$^{-1}$ is higher than the threshold at which pile-up becomes significant (${\approx}0.3$~counts~frame$^{-1}$). We use the {\tt jdpileup} model in {\sl Sherpa}, fitting the grade migration parameter $\alpha$ and the pile-up strength parameter $f$ (\citealt{davis2001}). We call the entire collection of spectral fits  the disentangled analysis.

For comparison, we also carry out a  spectral analysis of the sources based on a \naive\ allocation of photons that collects events from within $1''$ of the fitted location of each source and assumes that there is no contamination from the other source. The only difference in the spectral model for the \naive\ and disentangled analyses is in how the effective areas are defined. In the case of the \naive\ analysis, a correction is made post-facto to the normalization based on how much of the source is expected to be included within the $1''$ source photons extraction radius. In the disentangled analysis, the assumed extraction radius for the spectra with allocated events is set to be $2.5''$ and the subsequent correction is negligible.

The results of the spectral fits to the disentangled spectra are shown as histograms of best-fit values for $N_{\rm H}$, $kT$, $Z$, and model flux computed for each of the 1000 spectra, see Figure \ref{fig:secondart_analysis}. In several cases, a bimodal distribution is apparent. This suggests that a multi-temperature component spectrum would be a better fit. The separation of the modes, however, is generally too small to be picked up by typical multi-temperature model fits. Not shown are the pile-up parameters for the bright source, which are consistent between the \naive\ and disentangled analyses ($(\alpha,f)=(0.6,0.93)$ for \naive, and $(0.53,0.89)$ for the disentangled spectra), though the former indicates that the pile-up strength is slightly higher. This is to be expected, since the \naive\ analysis is carried out for photons in the core of the PSF, where naturally pile-up is most significant. The disentangled spectra include photons from the wings, thus reducing the strength of pile-up effects and decreasing the correction needed to the source flux by about $60$\%.

The spread in the histograms in Figure \ref{fig:secondart_analysis} indicates the uncertainty in the best-fit values due to uncertainty in the allocation of photons. The best-fit values from the \naive\ calculation are shown as solid red vertical lines. The dashed red vertical lines give 68\% intervals indicating the statistical errors, due to randomness in the photons emitted and detected, under the \naive\ analysis. These statistical errors do not account for uncertainty in the photon allocations. The histograms, on the other hand, represent only errors due to uncertainty in the photon allocations, but do not account for statistical errors (due to randomness in photon emission and detection). Because the two sources of error are independent, and because we expect the statistical errors for the disentangled analyses to be similar to those for the \naive\ analysis, the total errors could be represented by a perturbation of the histograms with $\sigma$ equal to the statistical errors from the \naive\ analysis. For these data, with the exception of flux (panels (g) and (h) of Figure \ref{fig:secondart_analysis}), the statistical errors dominate the errors due to uncertainty in the photon allocation.  Despite this, the disentangled analysis provides reasonable evidence that the absorption column of the faint source (panels (b) of Figure \ref{fig:secondart_analysis}) and the flux of the two sources (panels (g) and (h) of Figure \ref{fig:secondart_analysis}) are different from the  best-fit values under the \naive\ analysis.

The variability of the true parameters around each of the best fit values recorded in the histograms is expected to be similar to that indicated for the \naive\ fit. However, we did not calculate these uncertainties because of the large computational cost. For these data, with the exception of flux (panels (g) and (h) of Figure \ref{fig:secondart_analysis}), the variability in the true spectral parameters around the best fit values is likely larger than the uncertainty in the best fit values (due to the uncertainty in the allocation of photons). Despite this, the disentangled analysis provides reasonable evidence that the absorption column of the faint source (panel (b) of Figure \ref{fig:secondart_analysis}) and the flux of the two sources (panels (g) and (h) of Figure \ref{fig:secondart_analysis}) are different to the \naive\ analysis best-fit values.

Overall, the \naive\ analysis best-fit values for the fainter source are in greater disagreement with the disentangled analysis than those for the bright source.  This is to be expected, since in the \naive\ analysis, the contamination of the fainter source by the brighter source is larger. Our algorithm effectively removes this contamination. This causes the spectral fit parameter values to change and the measured source flux of the fainter source to decrease. In summary, the observed changes to the spectral model parameters are as would be expected when contamination is reduced and the data quality is improved.

\section{Summary}
\label{sec:summary}

We have developed a Bayesian statistical method that models spatial and spectral information from overlapping sources and the background, and jointly estimates all individual source parameters. The key contributions of our approach are the use of spectral information to improve spatial separation, coherent quantification of uncertainty, including that of the number of sources, and the probabilistic assignment of photons to the different sources. Our simulation studies show that using spectral information improves the detection of both faint and closely overlapping sources and increases the accuracy with which source parameters are inferred.

We have analyzed data from two sets of overlapping sources observed with XMM and {\sl Chandra}. Traditional analysis of XMM observations of FK and FL Aqr, thought to be a visual binary, show that their spectra are not distinguishable. Our analysis confirms that the spectra are indeed similar, but nonetheless shows that they are separable. We have also carried out detailed spectral analysis on disentangled photons from a pair of close sources from near the center of the Orion Nebula Cluster observed with {\sl Chandra}. We find that the spectral parameters change significantly after contamination is removed.

The data we have considered consists of event-level observations. In the more usual case of spatially binned data, the PSF could be updated to take account of the binning. If the spatial pixels are larger, the importance of spectral data is greater, because it is harder to spatially distinguish sources from each other and the background. Clearly however, unbinned data is preferred when available, and our method has the ability to use all the information in such data. Similar comments apply when the spectral data are grouped.

As with other detection procedures, an important question is how to combine information from multiple observations. Since our approach gives the posterior distribution of all the parameters, this can be used as the prior distribution is subsequent analyses. Thus, under the Bayesian framework it is straightforward to analyze the available observations sequentially, which is convenient in that different PSFs, for example, can be used for each analysis. This is critical if the observations are recorded by different observatories.

Another advantage of the Bayesian framework is that more complex models can straightforwardly be built in. For example, using a location or spectral dependent PSF would require only minimal changes to the method and code. Another extension is to include the different temporal signatures of overlapping sources to further separate them. Future work will focus on these and related issues as well as computational scalability.

\section*{Acknowledgements}

This work was supported by the Smithsonian Competitive Grants Program for Science Fund 40488100HH0043 and was conducted under the auspices of the CHASC International Astrostatistics Center. CHASC is supported by NSF grants DMS 1208791 and DMS 1209232. DEJ acknowledges support from the Harvard Statistics Department, VLK from a NASA contract to the {\sl Chandra} X-Ray Center NAS8-03060 and from Chandra grant AR0-11001X, and DvD from a Wolfson Research Merit Award provided by the British Royal Society and from a Marie-Curie Career Integration Grant provided by the European Commission. In addition, we thank CHASC members for many helpful discussions, especially Xiao-Li Meng, Andreas Zezas, Aneta Siemiginowska, Lazhi Wang, and Alex Blocker.

\section*{Appendices}

\subsection*{A. Split and combine proposals in reversible jump MCMC}

The purpose of this appendix is to detail our implementation of split-combine moves in {\coledits the BASCS code}. We assume the reader is familiar with MCMC and RJCMC algorithms. Those unfamiliar with MCMC we refer to \citet{textbook} and the appendix of \citet{xu2014}. Those unfamiliar with RJMCMC we refer to \citet{richard} and \citet{green}. The basic properties of the algorithm follow from the reversibility condition and the theory of Markov chain convergence dealt with in many probability and stochastic processes books, for example \citet{feller1968}.

We concentrate on the split proposals used in {\coledits BASCS} because they are more complex than the combine proposals. In particular, we detail the steps of a split proposal in {\coledits BASCS} for the extended full model (the most complex case considered). The corresponding combine proposals are straightforwardly obtained by solving the equations appearing in our split proposal scheme for the parameters of the combined source (i.e., the parameters of the yet to split source). Conditions that are required of newly split sources must also be satisfied when sources are combined. Following the algorithm is a short description of the reasons that its novel features are necessary in the current context.

Let $\p_j=(\p_{jx},\p_{jy})$ be the location of the source the algorithm is attempting to split. Throughout this appendix, the parameters for the two newly proposed sources formed by a split will be subscripted as in the main parts of the paper except that a 1 will appear after the subscript $j$ to indicate the first newly proposed source, and similarly a $2$ will indicate the second newly proposed source e.g. $\p_{j1\datx}$ will denote the $x$-coordinate of the first newly proposed source formed by a split. The newly proposed sources are ordered so that $\min(\gamma_{j11},\gamma_{j12})\leq\min(\gamma_{j21},\gamma_{j22})$, i.e., the smallest {\sl gamma} distribution mean of the spectral model for the first newly proposed source is smaller than that of the second newly proposed source. For the full model the ordering used is $\gamma_{j1}\leq\gamma_{j2}$, and for the spatial-only model it is $\p_{j1x}\leq\p_{j2x}$. These orderings are solely for the purposes of proposals; the label switching problem is discussed separately in Appendix B. A split proposal is performed as follows:

\noindent{\bf Step 1:} Spectral parameters proposal: simulate $u \sim \mbox{Uniform}(0,1)$.
\begin{enumerate}[(a)]
\item If $u > 0.5$, simulate $u_1\sim \mbox{Beta}(2,2)$, $t,v_2,v_3 \sim \mbox{Uniform}(0,1)$ and $v_4,v_5 \sim \mbox{{\sl gamma}}(5,5)$. For $a=\ewt_j/u_1$ and $b=(\ewt_j+u_1-1)/u_1$ define
\begin{eqnarray*}
f(u_1,\ewt_j) &=& \begin{cases}a & \mbox{if } a < 1\\
1 + e^{\frac{10}{a}-10}\log\left(a\right) & \mbox{otherwise,}\end{cases}\\
g(u_1,\ewt_j) &=& \begin{cases}b & \mbox{if }b > 0\\
be^{10b}\log\left(b\right) & \mbox{otherwise.}\end{cases}
\end{eqnarray*}
Then set
    \begin{eqnarray*}
\ewt_{j1} &=& tg(u_1,\ewt_j) + (1-t)h(u_1,\ewt_j)\\
\ewt_{j2} &=& \frac{\ewt_j-u_1\ewt_{j1}}{1-u_1}\\
\gm_{j11}  &=&v_{2}\gm_{j1}\\
\gm_{j21}  &=& \frac{1-v_{11}v_2}{1-v_{11}}\gm_{j1} \\
\gm_{j12}  &=&\gm_{j11}+\frac{v_3}{v_{12}}(\gm_{j2}-\gm_{j11})\\
\gm_{j22}  &=&\gm_{j11}+\frac{1-v_3}{1-v_{12}}(\gm_{j2}-\gm_{j11})\\
\epa_{j11}  &=& v_4 \epa_{j1}\\
\epa_{j12}  &=& v_5 \epa_{j2}\\
\epa_{j2l}  &=&  \frac{\ew_{j2l}\gm_{j2l}^2}{A_{jl}}\hspace{0.5cm}\mbox{for $l=1,2$},
\end{eqnarray*}
where
\begin{eqnarray*}
A_{jl} &=& \ew_{j1}\gm_{jl}^2\left(1+\frac{1}{\epa_{jl}}\right) \\
&-& \ew_{j1l}\gm_{j1l}^2\left(1+\frac{1}{\epa_{j1l}}\right)-\ew_{j2l}\gm_{j2l}^2,
\end{eqnarray*}
and $\ew_{j1}=\w_{j}\ewt_j$, $\ew_{j11}=\w_{j1}\ewt_{j1}$, $\ew_{j12}=\w_{j2}\ewt_{j2}$,  $\ew_{j2}=\w_j(1-\ewt_j)$, $\ew_{j12}=\w_{j1}(1-\ewt_{j1})$, and $\ew_{j22}=\w_{j2}(1-\ewt_{j2})$.
\vspace{0.2cm}

The split proposal is immediately rejected if $\ewt_{j1}$ is not between $\min\left(1,\frac{\ewt_{j}}{u_1}\right)$ and max
$\left(0,\frac{\ewt_j+u_1-1}{u_1}\right)$, or $\gm_{j21} > \gm_{j22}$, or any of $\gm_{j11}$, $\gm_{j21}$, $\gm_{j12}$, $\gm_{j22}$ are outside the range of the spectral data $\enc$.
\item If $u \leq 0.5$, simulate $\ewt_{j1},\ewt_{j2} \sim \mbox{Beta}(10,1)$, $v_2, v_3 \sim \mbox{Uniform}(0,1)$, and $v_4, v_5 \sim \mbox{Beta}(1,5)$. Then set
 $u_1 = \ewt_j$, $\gm_{j11} =\gm_{j1}$, $\gm_{j21} = \gm_{j2}$, $\epa_{j11} = \epa_{j1}$, $\epa_{j21} = \epa_{j2}$, and
 \begin{eqnarray*}
 \gm_{j12} &=& \gm_{j1} + v_2(\emax-\gm_{j1})\\
\gm_{j22} &=&\gm_{j2} + v_3(\emax-\gm_{j2})\\
\epa_{j12}&=& 20v_4\\
\epa_{j22} &=& 20 v_5.
 \end{eqnarray*}
\end{enumerate}

\noindent{\bf Step 2:} Spatial parameters proposal: simulate $u_1\sim \mbox{Beta}(2,2)$, $u_2\sim S_2\mbox{Beta}(2,2)$, and $u_3 \sim S_3\mbox{Beta}(2,2)$ (where $S_2$ and $S_3$ are independent random signs) and set
\begin{eqnarray*}
\w_{j1} &=& \w_ju_1 \\
\w_{j2} &=& \w_j(1-u_1)\\
\p_{j11} &=& \p_{j\datx}-u_2\sig\sqrt{\frac{\w_{j2}}{\w_{j1}}}\\
\p_{j21} &=& \p_{j\datx}+u_2\sig\sqrt{\frac{\w_{j1}}{\w_{j2}}}\\
\p_{j12} &=& \p_{j\daty}-u_3\sig\sqrt{\frac{\w_{j2}}{\w_{j1}}}\\
\p_{j22} &=& \p_{j\daty}+u_3\sig\sqrt{\frac{\w_{j1}}{\w_{j2}}}.
\end{eqnarray*}
In our algorithm $\sig=1$ (tuning parameter).
\vspace{0.5cm}

\noindent{\bf Step 3:} If $(\p_{j1\datx}-\p_{j'\datx})^2 + (\p_{j1\daty}-\p_{j'\daty})^2 < (\p_{j1\datx}-\p_{j2\datx})^2 + (\p_{j1\daty}-\p_{j2\daty})^2$, for some $j' \in \{1,\dots,\nk\}/\{j\}$, then the split proposal is rejected. We also reject the split proposal if the proposed source locations are outside the convex hull of the spatial data $(\datx,\daty)$.
\vspace{0.5cm}

\noindent{\bf Step 4:} To update $s$ to $s'$ randomly assign photon $i$ to the first newly proposed source with probability $p_i = p_{i1}/(p_{i1}+p_{i2})$, and otherwise to the second newly proposed source, for each $i \in \mathcal{I}_j$. Here
\begin{eqnarray*}
p_{il} = w_{jl}\fpsf_{(\p_{jl1},\p_{jl2})}(\datx_i,\daty_i)\sum_{r=1}^2\ewt_{jlr} \fspec_{\epa_{\alloc_{jlr}},\gm_{\alloc_{jlr}}}(\en_i),
\end{eqnarray*}
for $l=1,2$. We denote the probability of the particular allocation realized by $P_{alloc}$.
\vspace{0.5cm}

\noindent{\bf Step 5:} Simulate $u_{\mbox{\footnotesize{split}}}\sim \mbox{Uniform}(0,1)$ and accept the proposed split if  $u_{\mbox{\footnotesize{split}}} < \min\{1,A\}$ where $A$ is
 \begin{eqnarray*}
\begin{cases}\frac{p(\cparas_{\nk+1}',\nk+1,\alloc'|\datx,\daty,\en)}{p(\cparas_\nk,\nk,\alloc|\datx,\daty,\en)}\frac{d_{\nk+1}}{b_\nk P_{alloc}}\\
\times\frac{1}{\frac{1}{4}b_{2,2}(u_1)b_{2,2}(|u_2|)b_{2,2}(|u_3|)}\\
\times\frac{1}{g_{5,5}(v_4)g_{5,5}(v_5)}|J_a| \qquad \mbox{if $u > 0.5$ (Step 1)}\\\\
\frac{p(\cparas_\nk',\nk+1,\alloc'|\datx,\daty,\en)}{p(\cparas_\nk,\nk,\alloc|\datx,\daty,\en)}\frac{d_{\nk+1}}{b_\nk P_{alloc}}\\
\times\frac{1}{\frac{1}{4}b_{2,2}(|u_2|)b_{2,2}(|u_3|)b_{10,1}(\ewt_{j1})b_{10,1}(\ewt_{j2})}\\
\times\frac{1}{b_{1,5}(v_4)b_{1,5}(v_5)}|J_b| \qquad \mbox{otherwise.}\end{cases}\label{eqn:split_accept}
\end{eqnarray*}
Here, the notation $b_{S,R}$ and $g_{S,R}$ denotes the $\mbox{Beta}(S,R)$ and $\mbox{{\sl gamma}}(S,R)$ densities, respectively, and
\begin{eqnarray*}b_\nk = \begin{cases} \frac{1}{\nk} & \mbox{if }\nk=1\\ \frac{1}{2}\frac{1}{\nk}& \mbox{otherwise,}\end{cases}\qquad\qquad\qquad\qquad\\
d_{\nk+1} = \begin{cases} \frac{1}{\nk+1} & \mbox{if }||(\p_{j1\datx},\p_{j1\daty})-(\p_{j2\datx},\p_{j2\daty})||_2 \\ &\leq||(\p_{j1\datx},\p_{j1\daty})-(\p_{j'\datx},\p_{j'\daty})||_2,\\
& \forall j' \in \{1,\dots,\nk\}/\{j\}\\
 \frac{1}{2}\frac{1}{\nk+1} & \mbox{otherwise.}\end{cases}\end{eqnarray*}
The Jacobian $|J_a|$ is the determinant of a $16\times 16$ block matrix. The determinant of the upper-left $6 \times 6$ block is $\w_j\sig^2/(u_1(1-u_1))$, and this is multiplied by the determinant of the lower-right block which is calculated numerically.  The Jacobian $|J_b|$ is $20^2(\emax-\gm_{j1})(\emax-\gm_{j2})\w_j\sig^2/(u_1(1-u_1))$.
\vspace{0.2cm}

\begin{figure*}[t]
\includegraphics[trim = 0mm 0mm 0mm 20mm,width=1\textwidth]{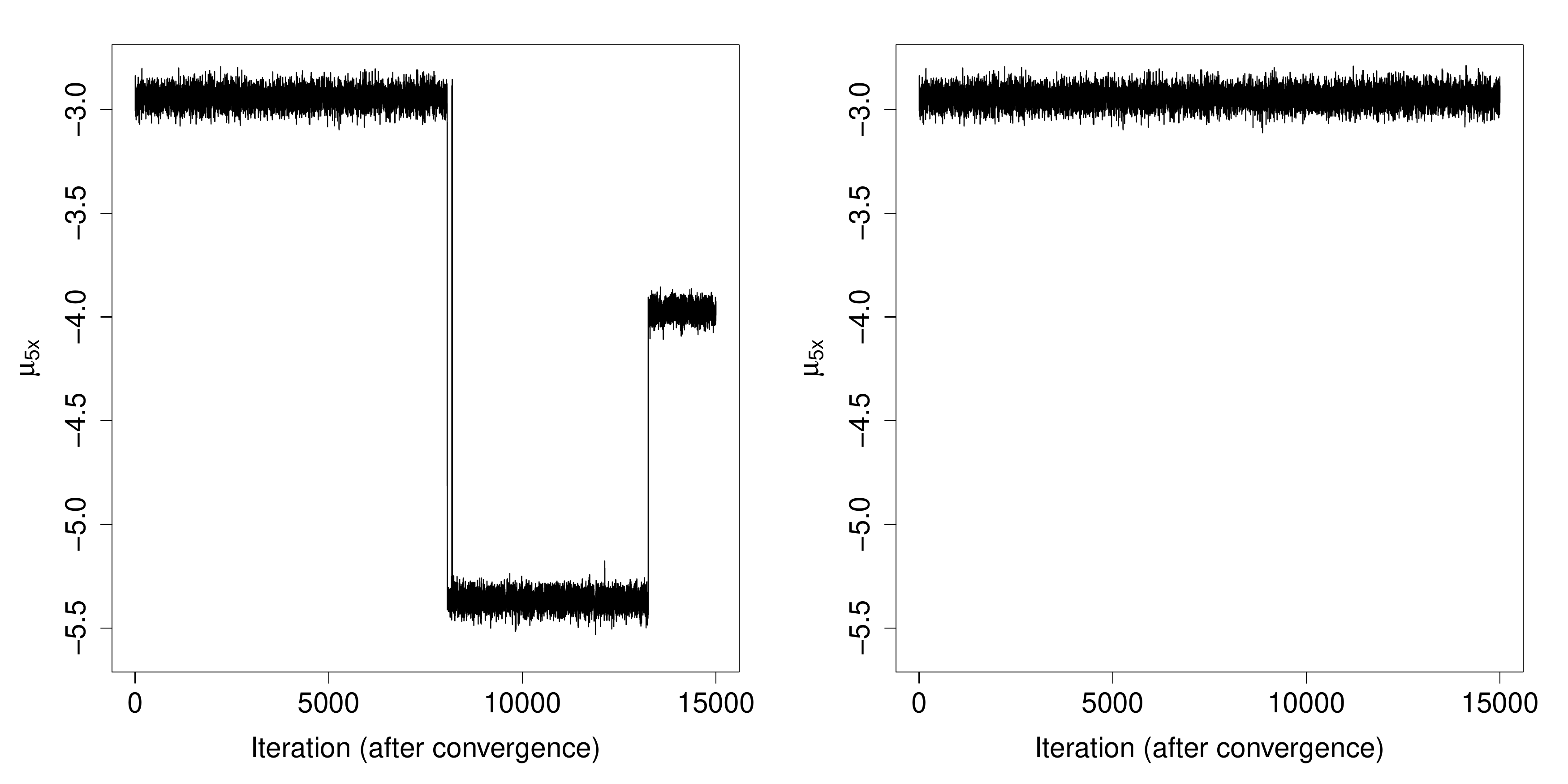}
\caption{Trace plot of the parameter $\mu_{5x}$ from a simulation with ten sources (Section \ref{sec:ksense}) before (left) and after (right) relabelling.\label{fig:label_switching}}
\end{figure*}

There are two features of {\coledits BASCS} that are not explicitly dealt with in standard approaches. The first is that the distributions we split and combine are themselves mixture distributions. The second is that {\coledits BASCS} randomly chooses from two proposal schemes for the spectral parameters in Step 1 because a single approach does not address all the possibilities. The approach in Step 1(a) splits each {\sl gamma} distribution in the current source's spectral model into two, thus forming two new spectral models for the newly proposed sources. The key aspect of this approach is that the new spectral models are designed to both be similar to the original. This makes sense in a situation where two similar sources have been mistaken for one. The approach in Step 1(b) is designed to split one true source into two, with each newly proposed source accounting for one {\sl gamma} component of the true spectral model. Thus, the two new source spectral models each typically have nearly all their weight on a single {\sl gamma}, which is almost invariably the first component in the extended full spectral model (we sort the {\sl gammas} by their means, in increasing order). Of course, we do not want to split a true source, but this split proposal is necessary in order to allow the reverse combine proposal, because the reversibility condition must be satisfied.

\subsection*{B. Label switching}
\label{app:label_switching}

A computational challenge is that the enumeration, or labelling, of individual sources changes stochastically during the iterations of an RJMCMC algorithm (and even during the iterations of an MCMC algorithm for a mixture model with a known number of components). For example, Figure \ref{fig:label_switching} shows the value of $\p_{5x}$ at each iteration of our algorithm (after convergence) before and after the labelling has been corrected (the data are from the simulation study involving ten sources described in Section \ref{sec:ksense}, and in particular, $\p_{5x}$ is the $x$-coordinate of the fifth source). Clearly, some such correction will be necessary in order for estimates such as that in Equation \ref{eqn:w2k_estimate} to be meaningful.

\begin{figure*}[t]
\includegraphics[trim = 0mm 0mm 0mm 35mm,width=1\textwidth]{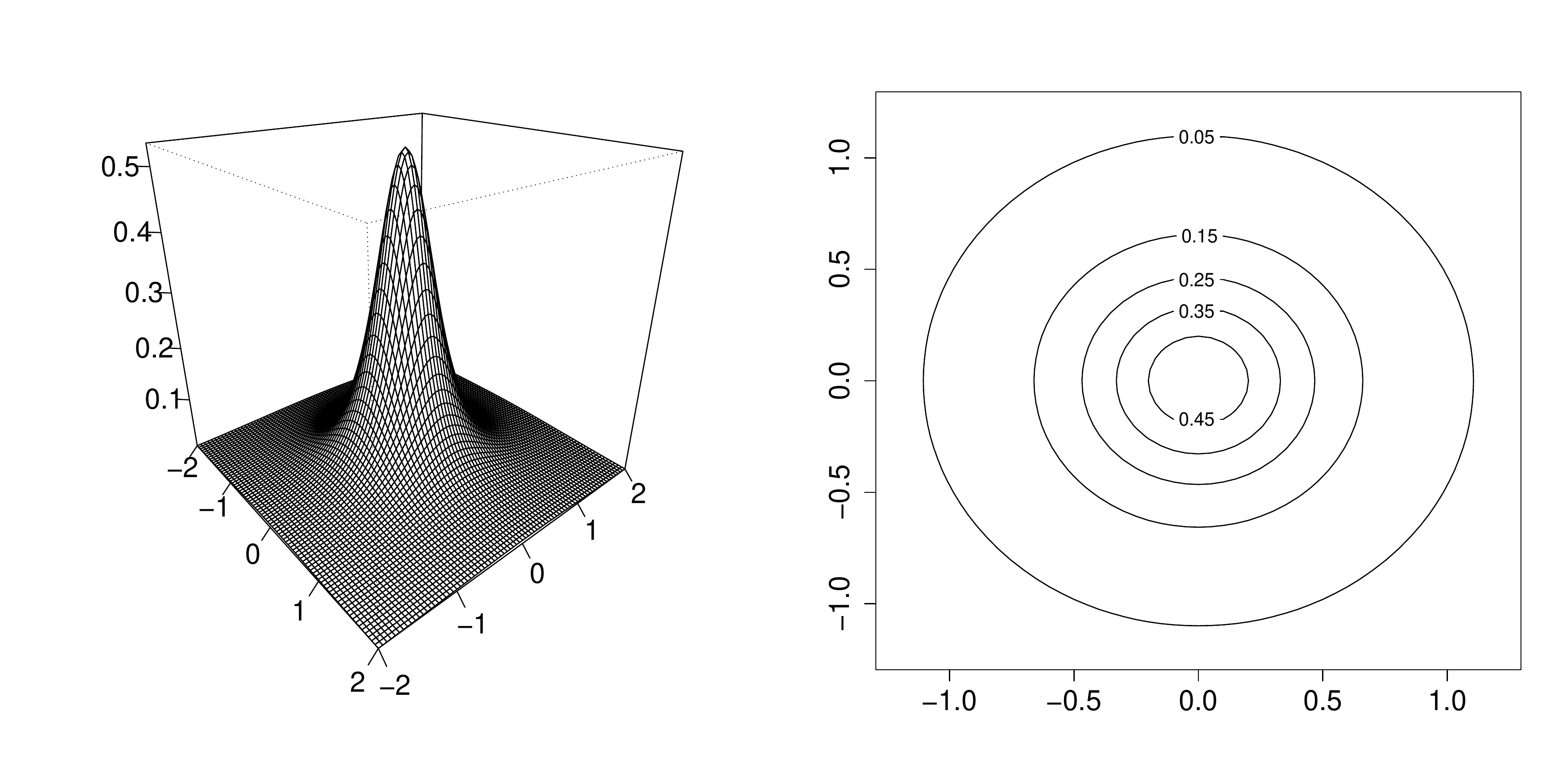}
\caption{2-D King profile density (left), and its contours (right).\label{fig:king}}
\end{figure*}

We implemented two approaches to relabelling and, in our real data analyses, they gave essentially identical results. The first method was to impose a hard constraint. In one dimension a hard constraint typically involves ordering the component locations, but it is not clear how best to impose such a constraint in two dimensions. As most of the source positions were precisely fitted, we simply ran the RJMCMC algorithm until convergence and then selected a posterior draw of the positions and weights to use as a reference. Running the algorithm again (or continuing the initial run), at each iteration we labelled the current source closest to the brightest reference source as source one, then we looked for the source closest to the second brightest reference source, and so on. As in the one dimensional case, this approach has the limitation that artificial `boundary' effects may be introduced when the posteriors of two source positions overlap. These effects indicate that the real posterior uncertainty has not been correctly recovered (unless there is some real information to support a hard constraint in our prior). However, in our real data analyses there was no evidence of such boundaries because, for probable values of $\nk$, all the source positions were precisely fitted and there was little overlap between the posteriors of source positions. In the case of the {\sl Chandra} observation and $\nk=14$, the fact that the posteriors of the source locations are non-overlapping can be seen from Figure \ref{fig:chandra_positions}.

We also implemented the approach suggested by \citet{west}, by modifying their publicly available code to work for our model. This method also uses a reference and is based on a loss function. At iteration $t$, the most likely assignment of each photon is computed treating the current parameter values as the true parameters, and then again treating the reference parameters as the true parameters. If, assuming the current parameter values, photon $i$ is most likely to have originated from component two, but another origin is most likely when assuming the reference parameter values, then we say there is a mismatch in allocation of photon $i$. The method used by \citet{west} is to choose the relabeling that minimizes the number of mismatches at iteration $t$, and then proceed to the next iteration. This second approach is substantially more computationally expensive than the first. Therefore we use the first approach online and apply the second only if there are potential `boundary' effects  (neither method is effected by the initial labels and therefore no problems are caused by applying both).

\subsection*{C. King profile}

The functional form of the 2-D King profile is
\begin{eqnarray*}
f(\polarr) = \frac{C}{(1+(\polarr/\polarz)^2)^\kingp}
\end{eqnarray*}
where
\begin{eqnarray*}
\polarr(\cartx,\carty,\polart)=\qquad\qquad\qquad\qquad\qquad\qquad\qquad\qquad\\
\sqrt{(\cartx\cos \polart + \carty\sin \polart)^2+\frac{(\carty\cos \polart - \cartx\sin \polart)^2}{(1-\epsilon)^2}}.
\end{eqnarray*}
The constant $C$ is determined numerically. The particular parameters we use for the 2-D King profile are  as follows; off-axis angle $\theta$ = 0 arcmin, core radius $d_0$ = 0.6 arcsec, power-law slope $\kingp$ = 1.5, ellipticity $\epsilon$ = 0.00574. The resulting probability density is displayed in Figure \ref{fig:king}.

\bibliographystyle{apj}

\end{document}